\documentclass[aps,prb,floatfix,twocolumn,superscriptaddress,hyperref=pdftex]{revtex4-2}

\usepackage{amssymb}
\usepackage{hyperref}
\hypersetup{
    colorlinks,
    linkcolor={blue!80!black},
    citecolor={blue!80!black},
    urlcolor={blue!80!black}
}
\usepackage{graphicx}
\usepackage{placeins}
\usepackage{booktabs}

\usepackage{upgreek}
\usepackage{siunitx}
\usepackage{physics}
\usepackage{soul}
\usepackage{xcolor}
\usepackage{tabularx,multirow,array}
\usepackage{siunitx}
\usepackage{chemmacros}
\usepackage[normalem]{ulem}

\newcommand{\ten}[1]{\boldsymbol{\mathrm{#1}}}

\begin{document}
\title{Effective conduction-band model for zincblende III-V semiconductors in the presence of strain: tuning the properties of bulk crystals and nanostructures}%\title{Tuning the conduction band properties of zincblende III-V semiconductors through strain in bulk crystals and nanostructures}
%\title{Tuning the conduction band properties of zincblende III-V semiconductors: strain-induced effects in bulk crystals and nanostructures}
%\title{Strain effects in the conduction band of zincblende III-V semiconductor compounds: bulk crystals and nanostructures}
\author{Samuel D. Escribano}
\email[Corresponding author: ]{samuel.diazes@gmail.es}
\affiliation{Department of Condensed Matter Physics, Weizmann Institute of Science, Rehovot 7610, Israel}
\author{Alfredo Levy Yeyati}
\affiliation{Departamento de Física Teórica de la Materia Condensada C5, Condensed Matter Physics Center (IFIMAC) and Instituto Nicolás Cabrera, Universidad Autónoma de Madrid, E-28049 Madrid, Spain}
\author{Elsa Prada}
\affiliation{Instituto de Ciencia de Materiales de Madrid (ICMM), Consejo Superior de Investigaciones Científicas (CSIC), E-28049 Madrid, Spain}

%\date{\today}
\begin{abstract}
%Strain provides a powerful knob to tailor the electronic properties of semiconductors. Simple yet accurate approximations that capture strain effects in demanding simulations of mesoscopic nanostructures are therefore highly desirable. However, for III–V compounds—key materials for quantum applications—such approaches remain comparatively underdeveloped. In this work, we derive a compact, effective Hamiltoniant hat incorporates strain effects into the conduction band of zincblende III–V semiconductors. Starting from the eight-band k$\cdot$p model with Bir–Pikus corrections, we fold down the valence bands to obtain analytical expressions for strain-renormalized parameters, including the effective mass, chemical potential, spin–orbit coupling, and $g$-factor. The model reproduces full multiband results under small to moderate strain, while retaining a form suitable for device-scale calculations. We benchmark the model for bulk deformations and apply it to representative nanostructures, such as core–shell nanowires and planar heterostructures. Our results provide a practical and versatile tool for incorporating strain into the design of III–V semiconductor devices, enabling reliable predictions of their properties with direct implications for spintronic, straintronic, optoelectronic, and topological quantum technologies.

Strain provides a powerful knob to tailor the electronic properties of semiconductors. Simple yet accurate approximations that capture strain effects in demanding simulations of mesoscopic nanostructures are therefore highly desirable. However, for III–V compounds—key materials for quantum applications—such approaches remain comparatively underdeveloped. In this work, we derive a compact, effective Hamiltonian that describes the conduction band of zincblende III–V semiconductors incorporating strain effects. Starting from the eight-band k$\cdot$p model with Bir–Pikus corrections, we perform a folding-down procedure to obtain analytical expressions for conduction-band strain-renormalized parameters, including the effective mass, chemical potential, spin–orbit coupling, and $g$-factor. The model reproduces full multiband results under small to moderate strain, while retaining a form suitable for device-scale calculations. We benchmark the model for bulk deformations and apply it to representative nanostructures, such as core–shell nanowires and planar heterostructures. Our results provide a practical and versatile tool for incorporating strain into the design of III–V semiconductor devices, enabling reliable predictions of their properties with direct implications for spintronic, straintronic, optoelectronic, and topological quantum technologies.

\end{abstract}

\maketitle

\section{Introduction}
The application of mechanical strain to low-dimensional semiconductors (SMs) has attracted growing interest in recent years due to its ability to modify and control the optoelectronic properties of these materials~\cite{Singh2003}. In particular, strain engineering in two-dimensional (2D) crystals has proven to be effective in tailoring band structures locally and globally~\cite{IOP:Roldan15, npj:Chaves20, LSA:Peng20, NPJ:Miao21, AdvMat:Kim23, AdvMat:Qi23, NanoLet:Vasconcelos25}. In contrast, strain manipulation in III–V SM compounds, which are widely used in both research and industry for a broad range of applications, remains comparatively less mature. To date, it has been demonstrated only in a narrow set of very specific platforms,
%~\cite{JAP:Jones17, NanoLet:18, NatCom:Balaghi19, NatCom:Balaghi21, PRB:Mondal23, PRL:Kato04, PRL:Chang07, PRB:Habib07, SciRep:Yang19, NanoLet:Skold05, NatCom:Signorello14, IOP:Alekseev15, NanoLet:Greil16, NanoLet:Hetzl16, Micromachines:Alekseev20, JVSTBV:Ahammou21, MicEng:Stomeo07, NRL:Li13, APL:Carter17, APL:Nguyen21}. In this kind of materials, strain is typically introduced, either intentionally or inadvertently, through lattice mismatch in carefully designed heterostructures, where an active layer is grown on top of insulating buffer layers with different lattice constants. This approach 
which has opened new avenues for potential applications in quantum technologies~\cite{JAP:Jones17, NanoLet:18, NatCom:Balaghi19, NatCom:Balaghi21, PRB:Mondal23}, including in straintronics~\cite{PRL:Kato04, PRL:Chang07, PRB:Habib07, SciRep:Yang19, PRB:Tholen16}, optoelectronics~\cite{NanoLet:Skold05, NatCom:Signorello14, IOP:Alekseev15, NanoLet:Greil16, NanoLet:Hetzl16, Micromachines:Alekseev20, JVSTBV:Ahammou21} and sensing~\cite{MicEng:Stomeo07, NRL:Li13, APL:Carter17, APL:Nguyen21}.

\begin{figure}[h!]
    \centering
    \includegraphics[width=1\columnwidth]{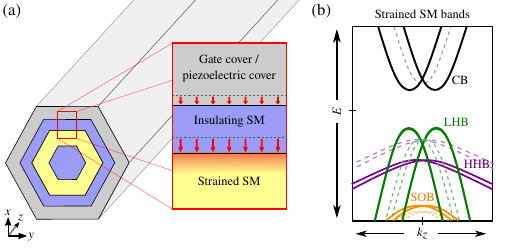}
    \caption{\textbf{Strain effects.} (a) Schematic of a strained core-double shell nanowire (NW). The inner core and outer shell (blue) are made of an insulating semiconductor (SM), which confines the electronic transport to the intermediate active shell (yellow), composed of a different SM material. Due to lattice mismatch between the layers, the active shell may experience strain, deforming from its original shape (dashed outline) to a modified one (solid outline). Alternatively, strain can be externally induced or tuned via a metallic piezoelectric gate (gray) that covers the entire nanostructure, which otherwise can simply be used as a potential gate. (b) Illustration of the effects of strain on the electronic band structure of the active SM layer. In III-V SM compounds, the low-energy spectrum consists of the spinful conduction band (CB) (black), light-hole band (LHB) (green), heavy-hole band (HHB) (purple), and split-off band (SOB) (orange). Strain shifts and distorts these bands (dashed: unstrained; solid: strained), modifying key parameters such as effective mass, chemical potential, spin-orbit coupling, and Land\'e $g$-factor.
    }
    \label{Fig0}
\end{figure}

At the same time, strain can also play an uncontrolled and often detrimental role in nanoscale devices, where it may hinder the expected performance. This is particularly relevant in heterostructures designed to host topological phases, such as topological insulators or topological superconductors, where strain may unintentionally break their topological protection~\cite{PRB:Gong13, PRL:Du17, IOP:Bjergfelt19, PRM:Ahn21}. Despite this, strain effects are frequently neglected in theoretical models~\footnote{Frequently, strain effects are  treated using multiband k$\cdot$p formulations~\cite{Bir1974, PRB:Bahder90} or fully \textit{ab initio} methods~\cite{PRB:Chantis08, PRB:Li21, PRB:Mondal23}, but these approaches, while accurate, are difficult to use for predicting or designing nanodevices with desired functionalities. Analytical treatments based on perturbation theory~\cite{Winkler2003} or symmetry arguments~\cite{PRB:Zemouri25} exist, yet they remain too general and, therefore, of limited and/or direct practical use.}, as the associated multiband contributions are complex and difficult to incorporate into the low-energy effective Hamiltonians typically used to describe such systems.

In this work, we address this problem by deriving comprehensive analytical expressions that capture the impact of strain on the conduction band (CB) of III–V SMs. Our main result is an effective CB Hamiltonian that accurately incorporates strain effects through renormalized parameters. The model is applicable to both bulk and low-dimensional systems and reproduces well the results obtained with more complex, multiband Hamiltonians for small to moderate strain. We explore various limiting cases of this model and examine how the effective parameters evolve with strain. Using this framework, we analyze representative one-dimensional (1D) and 2D nanostructures, demonstrating how strain can be harnessed to enhance desirable properties or, conversely, how its unintended presence may compromise device performance. 

A particularly promising architecture is the core-double shell nanowire (NW) sketched in Fig.~\ref{Fig0}(a). In this design, there is an insulating SM both at the inner core and the outer shell (in blue), thereby confining the electronic transport to the intermediate shell made of a distinct III–V compound (yellow). Because of the lattice mismatch between the layers, the active shell is naturally strained, and its band structure can be further tuned by an external piezoelectric gate that wraps the entire NW (gray). As we show in Section~\ref{Sec3}, such a geometry provides a versatile platform to control key electronic parameters through mechanical stress, specifically the spin–orbit (SO) coupling and effective $g$-factor, turning the NW into a functional straintronic nanodevice.

This work is structured as follows. In Section~\ref{Sec1}, we derive the analytical expression that describes the CB of III-V SM compounds under strain effects, and present the equations for all relevant effective parameters. Section~\ref{Sec2} examines how these parameters vary under different types of stress in a bulk SM. We focus on three fundamental deformation types: hydrostatic (Section~\ref{Sec2-1}), biaxial deviatoric (Section~\ref{Sec2-2}), and pure shear (Section~\ref{Sec2-3}). Notice that any deformation, excluding a rotation, can be expressed as a linear combination of these. In Section~\ref{Sec3}, we shift our attention to specific low-dimensional nanodevices (1D and 2D), analyzing the impact of strain on their characteristic parameters. We also propose methods to detect strain within these nanodevices and suggest designs whose properties can be optimized through targeted stress engineering or straintronics. Finally, we conclude in Section~\ref{Sec4}. For a comprehensive summary of the obtained effective parameters across the entire III–V SM family, we refer the reader to Appendix~\ref{AP:other}.

\section{Strained conduction-band Hamiltonian}
\label{Sec1}

The low-energy spectrum of III-V SM compounds can be well described with the Kane 8-band (8B) k$\cdot$p model~\cite{JPCS:Kane57, Winkler2003}. This model is widely used in the literature and has become a reliable method to obtain the electronic properties of III-V SM compounds of both bulk and low-dimensional systems. Within this approach, the Hamiltonian describing the SM only involves the four lowest energy bands (eight taking into account the spin), see Fig.~\ref{Fig0}(b), written in the envelope function approximation (k$\cdot$p theory). Their interactions with higher-energy bands are treated perturbatively using L\"owdin perturbation theory. These perturbative elements are substituted by constant parameters that modify the coupling among the main bands. These parameters can be extracted from \textit{ab initio} calculations or experiments~\cite{JAP:Vurgaftman01}, making this model reliable.

Strain effects can be taken into account directly in the 8B Hamiltonian following the approach of Ref.~\onlinecite{PRB:Bahder90}. The strain deforms the original unit cell of the crystal lattice, described by the cell constants $\vec{c}=(c_x,c_y,c_z)$, to a different \emph{strained} unit cell with lattice constants
\begin{equation}
    \vec{c}_{\rm s}=(\ten{1}+\ten{e})\vec{c},
\label{Eq:c_strain}
\end{equation}
where $\ten{1}$ is the $3\times3$ identity matrix and
\begin{equation}
    \ten{e}=\begin{pmatrix}
e_{xx} & e_{xy} & e_{xz}\\
e_{yx} & e_{yy} & e_{yz} \\
e_{zx} & e_{zy} & e_{zz}
\end{pmatrix},
\end{equation}
is the strain tensor, whose definition follows from Eq.~\eqref{Eq:c_strain},
\begin{equation}
    (\ten{e})_{ij}\equiv\frac{(\vec{c}_{\rm s})_i-(\vec{c})_{i}}{(\vec{c})_{j}}.
\end{equation}
Each element of this tensor $e_{ij}$ provides the relative change of the unit cell along the $i$ direction when a deformation is applied in the $j$ one. The diagonal elements are called \emph{normal} strains while the non-diagonal terms are called \emph{shear} strains.

The effect of the strain in the k$\cdot$p Hamiltonian is to perturb the electrostatic potential, which is divided into two contributions: $\phi(\vec{r})$ describes the slow varying part, typically on the mesoscopic scale; and the crystal potential $\phi_{\rm c}(\vec{r})$ describes the periodic potential due to the underlying crystal structure, typically below the nanoscale and integrated out in the Kane Hamiltonian. Since the strain changes the periodicity of the crystal structure, $\phi_{\rm c}(\vec{r})$ is the only one affected by it. If the strain $\ten{e}$ is small, one can expand the crystal electrostatic potential perturbatively (to first order), 
\begin{equation}
    \phi_{\rm c,\mathrm{s}}(\vec{r})\simeq \phi_{\rm c}(\vec{r})+\sum_{i,j}  e_{ij} \left(\frac{\partial \phi_{\mathrm{c,s}}(\vec{r})}{\partial e_{ij}}\right)_{e_{ij}=0},
\label{Eq:potential_strain}
\end{equation}
and introduce strain also as an additional perturbative term in the 8B k$\cdot$p Hamiltonian. The resulting strained 8B k$\cdot$p Hamiltonian $H_{\rm 8B,s}$, called the Bir-Pikus Hamiltonian and derived in Ref.~\onlinecite{Bir1974}, can be found in Appendix~\ref{AP:eqs}. Notice that one key assumption in the derivation of this Hamiltonian is that the strain is spatially independent (that is, constant in space). Otherwise, it would not be possible to define a crystal unit cell and, thus, apply k$\cdot$p theory.

The state basis of the 8B model is spanned by the four bands closest to the SM gap, see Fig.~\ref{Fig0}(b), which comprise the conduction band (CB) and three valence bands (VBs), called the light-hole (LHB), heavy-hole (HHB) and split-off (SOB) bands. III-V binary compound SMs exhibit a direct gap around the $\Gamma$ point, with the Fermi level typically located within the gap and closer to the CB. It is possible to obtain an effective Hamiltonian for the CB by folding down the other bands, including them as a self-energy in the CB. Without strain, one can obtain an analytical expression~\cite{PRB:Darnhofer93, Winkler2003} by expanding this self-energy to first-order in perturbation theory~\cite{PRB:Wojcik18, PRR:Escribano20}, under the assumption that the coupling between the CB and VBs, denoted $P$, is much larger than the couplings between the VBs, $\gamma_i$. To the best of our knowledge, this folding of the bands has not yet been performed including strain effects.

Here, we perform a CB approximation using the approach described above but in the presence of strain. Apart from the assumptions mentioned above, we have to consider that the corrections to the couplings between VBs due to strain, $d_{\rm v}$, are also much smaller than the conduction-to-valence band couplings, $P$ and its correction $d_{\rm cv}$. Under these assumptions, we obtain a strained Hamiltonian, only valid for zincblende crystals, that we write as the sum of two terms
\begin{eqnarray}
H_{\rm CB,s} & = & H_{\rm CB,s}^{\rm (iso)} + H_{\rm CB,s}^{\rm (ani)}.
\label{Eq:CB_approx}
\end{eqnarray}
The isotropic term $H_{\rm CB,s}^{\rm (iso)}$ can be written compactly regardless of the momentum and spin direction, while the anisotropic term $H_{\rm CB,s}^{\rm (ani)}$ arises from the anisotropy of the zincblence crystal cell, and it modifies the properties along the (001) crystallographic direction ($z$ direction in our case). They are given by the expressions
\begin{eqnarray}
H_{\rm CB,s}^{\rm (iso)} & = &   \left(\vec{k}\cdot\frac{\hbar^2}{2\ten{m_{\rm s}^{\rm (iso)}}}\cdot\vec{k}^T+\mu_{\rm s}^{\rm (iso)} (\vec{r})-e\phi(\vec{r})\right)\sigma_0 \nonumber \\ 
&+&\frac{1}{2}\left[\vec{\alpha}_{\rm s}^{\rm (iso)}(\vec{r})\times \vec{k}-\vec{k}\times\vec{\alpha}_{\rm s}^{\rm (iso)}(\vec{r})\right]\cdot\vec{\sigma}^T \nonumber \\
&+& \frac{1}{2}\mu_{\rm B} \vec{B}\cdot \ten{g_{\rm s}^{\rm (iso)}}\cdot \vec{\sigma}^T, \\
H_{\rm CB,s}^{\rm (ani)} & = &  \left(\vec{k}\cdot\frac{\hbar^2}{2\ten{m_{\rm s}^{\rm (ani)}}(\vec{r})}\cdot\vec{k}^T+\mu_{\rm s}^{\rm (ani)} (\vec{r})\right)\sigma_0 \nonumber \\
&+&\frac{1}{2}\left[\vec{\alpha}_{\rm s}^{\rm (ani)}(\vec{r})\times \vec{k}-\vec{k}\times\vec{\alpha}_{\rm s}^{\rm (ani)}(\vec{r})\right]\cdot(0,0,\sigma_z)^T \nonumber \\
&+& \frac{1}{2}\mu_{\rm B} \vec{B}\cdot \ten{g_{\rm s}^{\rm (ani)}}\cdot (0,0,\sigma_z)^T ,
\end{eqnarray}
with $\vec{k}$ the momentum operator vector, and $\vec{\sigma}$ the Pauli matrices. Notice that both Hamiltonians have the same functional form but the anisotropic one contains corrections coming from the $z$ direction.

The first term in these equations is the kinetic energy, with $\ten{m_{\rm s}^{\rm (iso)}}$ and $\ten{m_{\rm s}^{\rm (ani)}}$ the effective isotropic and anisotropic masses, respectively. The second term, either $\mu_{\rm s}^{\rm (iso)} (\vec{r})-e\phi(\vec{r})$ or $\mu_{\rm s}^{\rm (ani)}(\vec{r})$, is the (slow-varying) electrochemical potential. The third term corresponds to the SO interaction, with $\vec{\alpha}_{\rm s}^{\rm (iso)}(\vec{r})$ the isotropic SO coupling and $\vec{\alpha}_{\rm s}^{\rm (ani)}(\vec{r})$ the anisotropic one. The fourth term is the Zeeman interaction, being $\ten{g_{\rm s}^{\rm (iso)}}$ and $\ten{g_{\rm s}^{\rm (ani)}}$ the effective isotropic and anisotropic $g$-factors, respectively.

The effective isotropic strained parameters are given by the expressions
\begin{equation}
\begin{aligned}
&\frac{1}{\ten{m_{\rm s}^{\rm (iso)}}} = \\ 
&\frac{1}{m_e}+ \frac{2P^2}{3\hbar^2}\left(\ten{1}-\ten{e}\right)^2\left(\frac{\sfrac{1}{2}}{\Delta_{\rm LHB, s}} + \frac{\sfrac{3}{2}}{\Delta_{\rm HHB, s}}+\frac{1}{\Delta_{\rm SOB, s}} \right),
 \end{aligned}
\end{equation}
\begin{equation}
\begin{aligned}
 & \mu_{\rm s}^{\rm (iso)} (\vec{r})=  \\
 &-\frac{q_{e}Pd_{\rm cv}}{3}\left(\frac{\sfrac{1}{2}}{\Delta_{\rm LHB, s}^2} + \frac{\sfrac{3}{2}}{\Delta_{\rm HHB, s}^2}+\frac{1}{\Delta_{\rm SOB, s}^2} \right)\vec{\nabla}\phi(\vec{r})\cdot\vec{e}_{\rm s}, 
 \end{aligned}
\end{equation}
\begin{equation}
\begin{aligned}
 \vec{\alpha}_{\rm s}^{\rm (iso)} (\vec{r}) & =  \frac{q_e P^2}{3}\mathrm{adj}\left\{\ten{1}-\ten{e}\right\}\left(\frac{1}{\Delta_{\rm LHB, s}^2}-\frac{1}{\Delta_{\rm SOB, s}^2} \right)\vec{\nabla}\phi(\vec{r})  \\
&-\frac{Pd_{\rm cv}}{3}\left(\frac{1}{\Delta_{\rm LHB, s}}-\frac{1}{\Delta_{\rm SOB, s}} \right)\vec{e}_{\rm s},
 \end{aligned}
\end{equation}
\begin{equation}
 \ten{g_{\rm s}^{\rm (iso)}} = g_e\ten{1}-\frac{2}{3}\frac{2m_eP^2}{\hbar^2}\mathrm{adj}\left\{\ten{1}-\ten{e}\right\} \left(\frac{1}{\Delta_{\rm LHB, s}}-\frac{1}{\Delta_{\rm SOB, s}} \right),
\end{equation}
with $q_e$, $m_e$ and $g_e$ the free electron's charge, mass and $g$-factor, respectively. $P$ and $d_{\rm cv}$ are the Kane parameters corresponding to the unstrained conduction-to-valence band coupling and its correction due to strain. The operator $\mathrm{adj}\left\{\cdot\right\}$ corresponds to the adjugate matrix operation, and $\mathbf{1}$ is the $3\times3$ identity tensor. The gaps $\Delta_{\rm LHB, s}$, $\Delta_{\rm HHB, s}$ and $\Delta_{\rm SOB, s}$ correspond to the strained LHB, HHB and SOB gaps, respectively, which are given by
\begin{eqnarray}
\Delta_{\rm LHB, s}=\Delta_{\rm g}-\left(a+\frac{b}{2}\right)\mathrm{tr}\left\{\ten{e}\right\}-\frac{3b}{2}e_{zz}-d_{\rm cv}\norm{\vec{e}_{\rm s}}, \label{Eq:LH_edge}\\
\Delta_{\rm HHB, s}=\Delta_{\rm g}-\left(a-\frac{b}{2}\right)\mathrm{tr}\left\{\ten{e}\right\}+\frac{3b}{2}e_{zz}+d_{\rm cv}\norm{\vec{e}_{\rm s}}, \label{Eq:HH_edge} \\
\Delta_{\rm SOB, s}=\Delta_{\rm g}+\Delta_{\rm soff}-\mathrm{tr}\left\{\ten{e}\right\},
\label{Eq:SOff_edge}
\end{eqnarray}
with $\Delta_{\rm g}$ and $\Delta_{\rm soff}$ the unstrained SM and split-off band gaps, $a$ and $b$ the deformation potential constants (i.e., another Kane parameters related to strain), and $\vec{e}_{\rm s}=(e_{yz},e_{zx},e_{xy})$ the shear strain vector. 

The anisotropic effective parameters are defined similarly as
\begin{equation}
\begin{aligned}
& \left(\frac{1}{\ten{m_{\rm s}^{\rm (ani)}}}\right)_{jk}  =   \\
& \frac{P^2}{\hbar^2}\left(\ten{1}-\ten{e}\right)_{j z}\left(\frac{1}{\Delta_{\rm LHB, s}}- \frac{1}{\Delta_{\rm HHB, s}} \right)\left(\ten{1}-\ten{e}\right)_{zk},
 \end{aligned}
\end{equation}
\begin{equation}
\begin{aligned}
& \mu_{\rm s}^{\rm (ani)} (\vec{r}) = \\
&-\frac{q_e Pd_{\rm cv}}{2}\left(\frac{1}{\Delta_{\rm LHB, s}^2}-\frac{1}{\Delta_{\rm HHB, s}^2} \right)\partial_z\phi(\vec{r})\cdot\left(\vec{e}_{\rm s}\right)_z,
 \end{aligned}
\end{equation}
\begin{equation}
\begin{aligned}
&\vec{\alpha}_{\rm s}^{\rm (ani)}(\vec{r}) =\\
&-\frac{q_e P^2}{2}\mathrm{adj}\left\{\ten{1}-\ten{e}\right\}\left(\frac{1}{\Delta_{\rm LHB, s}^2}-\frac{1}{\Delta_{\rm HHB, s}^2} \right)\vec{\nabla}\phi(\vec{r}) \\
&+\frac{Pd_{\rm cv}}{2}\left(\frac{1}{\Delta_{\rm LHB, s}}-\frac{1}{\Delta_{\rm HHB, s}} \right)\vec{e}_{\rm s},
 \end{aligned}
\end{equation}
\begin{equation}
\ten{g_{\rm s}^{\rm (ani)}}= \frac{2m_eP^2}{\hbar^2}\mathrm{adj}\left\{\ten{1}-\ten{e}\right\}\left(\frac{1}{\Delta_{\rm LHB, s}}- \frac{1}{\Delta_{\rm HHB, s}} \right).
\end{equation}
They all involve the difference between the inverse of the LHB and HHB edges, being zero if the LHB and HHB are degenerate.

In addition, when the magnetic field is nonzero, the momentum $\vec{k}$ must be understood as the \textit{canonical} momentum, i.e., $\vec{k}\rightarrow -i\vec{\nabla}-\frac{q_e}{\hbar}\vec{A}$, being $\vec{A}$ the magnetic potential vector. This introduces additional terms into the Hamiltonian~\cite{PRB:Czarnecki24} that renormalize the kinetic energy, the chemical potential, and the $g$-factor, increasing the complexity of the expressions. These additional interactions are known as (mesoscopic) orbital effects, as the magnetic potential vector is a spatially dependent parameter, and so are the renormalized parameters. For simplicity, we neglect these terms in this work, as their contribution strongly depends on the system analyzed.

It is worth noting that the effective Hamiltonian derived here does not contain any Dresselhaus SO terms. This is expected as our analysis is restricted to first-order L\"owdin perturbation theory within the multiband k$\cdot$p framework~\cite{Winkler2003}. At this level, the CB remains inversion-symmetric, and the only SO contribution that arises is of Rashba type, stemming from structural inversion asymmetry and the strain-induced modifications of the bands. Dresselhaus terms appear only at higher orders, through coupling to remote bands beyond the leading Kane model, and are therefore absent from our present formulation (even though they could become relevant at large strain constants). In practice, however, Dresselhaus couplings are generally much weaker than Rashba ones in III–V SM compounds~\cite{PRB:Campos18}.

\section{Bulk semiconductor}
\label{Sec2}

We consider three different types of strain applied to a bulk SM~\cite{Cleland2011}: \textit{i)} A uniaxial hydrostatic strain, which involves a deformation of the crystal potential along the same direction in which stress is applied. In this case, both the shape and the volume of the unit cell change. \textit{ii)} A biaxial deviatoric strain, which is a biaxial hydrostatic strain that when stress is applied in one direction, the unit cell stretches (or compresses) in that direction and simultaneously compresses (or stretches) in the perpendicular direction in order to preserve its volume. \textit{iii)} A pure shear strain, which is a deformation that preserves the volume and the size of the surface of the crystal unit cell without creating any rotation. Notice that any general deformation can be written as a linear combination of a hydrostatic and a pure shear deformation.

\subsection{Hydrostatic deformation}
\label{Sec2-1}
\begin{figure}
    \centering
    \includegraphics[width=1\columnwidth]{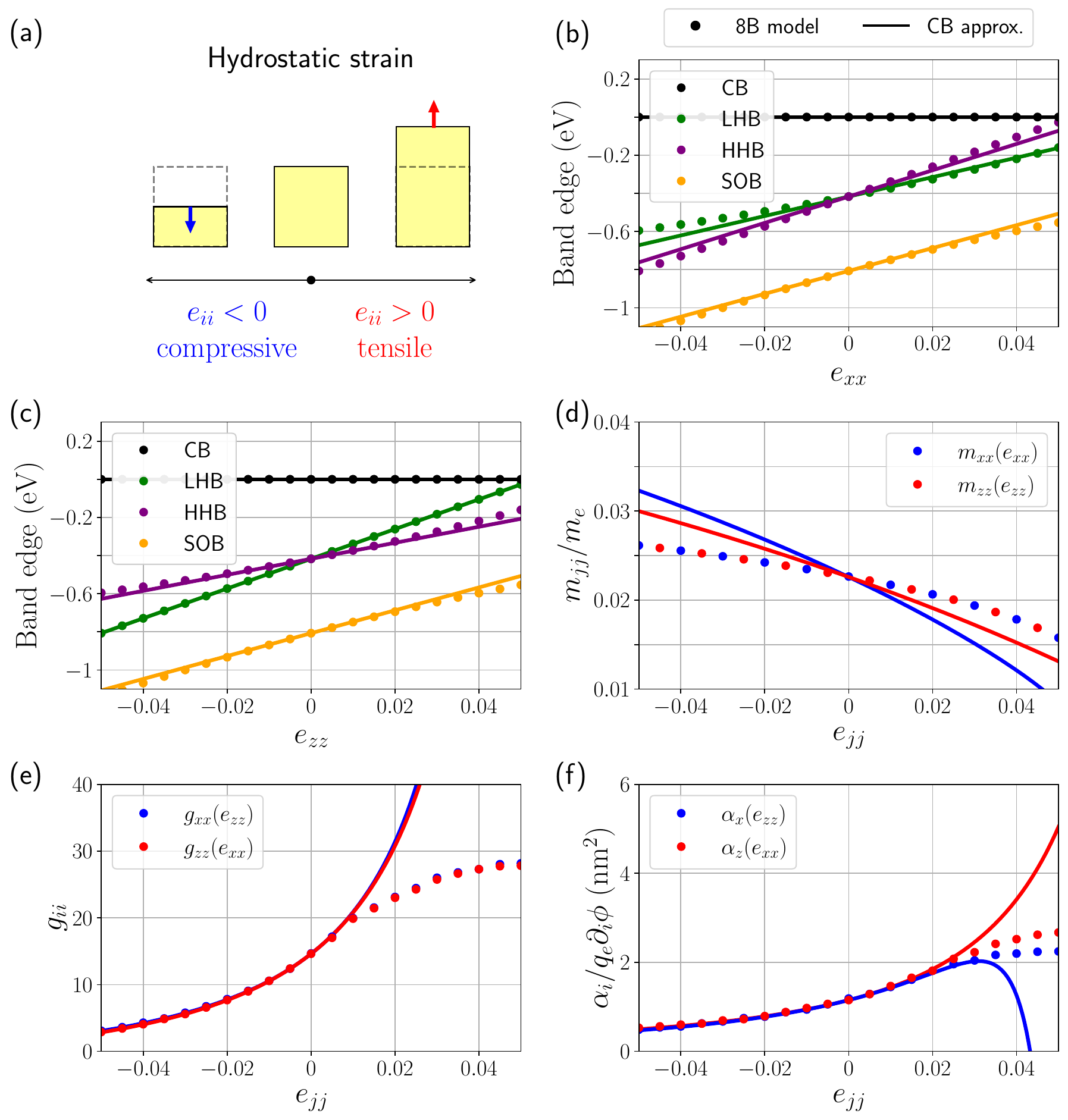}
    \caption{\textbf{Uniaxial hydrostatic strain.} (a) Sketch of an uniaxial hydrostatic deformation. A negative strain means a compressive deformation, while a positive one is a tensile deformation. (b) Band edge of the CB (black), LHB (green), HHB (purple) and SOB (orange) versus $e_{xx}$, i.e., when the hydrostatic strain is applied along the $x$ direction. In all the panels, dots represent the results of the full-diagonalization of the 8B model Hamiltonian, while solid lines represent the results provided by the CB approximation. (c) Same as (b) but along the $z$ direction. (d) Effective mass as a function of strain. We show $m_{xx}$ vs $e_{xx}$ in red, while $m_{zz}$ vs $e_{zz}$ in blue. (e) Effective $g$-factor vs strain. We show $g_{zz}$ vs $e_{xx}$ in red, while $g_{xx}$ vs $e_{zz}$ in blue. Notice that $g_{zz}$ ($g_{xx}$) does not depend on $e_{zz}$ ($e_{xx}$). (f) Spin-orbit (SO) coupling $\alpha$, normalized to the electric field $q_e\partial\phi$, as a function of strain. We show $\alpha_z$ vs $e_{xx}$ in blue, while $\alpha_x$ vs $e_{zz}$ in red. Similarly, $\alpha_z$ ($\alpha_x$) does not depend on $e_{zz}$ ($e_{xx}$). In all these simulations, we use parameters for bulk InAs, extracted from Ref.~\onlinecite{JAP:Vurgaftman01} and summarized in Appendix~\ref{AP:parameters}.
    }
    \label{Fig1}
\end{figure}

We start by considering the simplest case of a strain deformation: a compression or a stretch along one direction when stress is applied in that same direction; see Fig.~\ref{Fig1}(a). In this case, the strain tensor takes the form
\begin{equation}
    \ten{e}=\begin{pmatrix}
e_{xx} & 0 & 0\\
0 & e_{yy} & 0 \\
0 & 0 & e_{zz}
\end{pmatrix}.
\end{equation}
We analyze independently deformations along the $x$ and $z$ directions, as they are not crystallographic equivalent. We thus take $e_{yy}=0$ since the $y$ direction is equivalent to the $x$ one. In Fig.~\ref{Fig1}(b-f) we show how all the Hamiltonian parameters change with either $e_{xx}$ or $e_{zz}$. In all these plots, solid lines represent parameters computed with the CB approximation [Eq.~\eqref{Eq:CB_approx}], while dots represent parameters computed with the 8B model Hamiltonian (extracted by fitting the spectrum resulting from the diagonalization of the full  Hamiltonian; see Appendix~\ref{AP:num} for details).
%while dots represent parameters extracted by fitting the spectrum resulting from diagonalizing the full 8B model Hamiltonian (see Appendix for details). Let us dive on each parameter individually.

In Figs.~\ref{Fig1}(b,c) we show the behavior of the different band edges with $e_{xx}$ (b), when a deformation along $x$ is applied, and with $e_{zz}$ (c), when it is applied along the $z$ direction. We freely fix the Fermi level to the bottom of the CB. In both cases, a compressive strain widens the SM gap while a tensile strain narrows it. The LHB and HHB split with strain and their relative energy position depends on both the sign of the strain and the direction in which it is applied. This is a well-known fact and is due to the asymmetry of the crystal structure~\cite{Winkler2003}, which is especially enhanced when strain is applied. Notice that the CB approximation, compared to the 8B model results, successfully predicts the position of the band edges for $\left|e_{ii}\right|\lesssim0.05$, i.e., for small and moderate strains. Otherwise, the gap is too small and most likely the curvature of the bands is not properly accounted for.

The effective mass is shown in Fig.~\ref{Fig1}(d). In red (blue) we plot the $m_{zz}$ ($m_{xx}$) component when $e_{zz}$ ($e_{xx}$) is varied, being $\ten{m}^{-1}=(\ten{m_{\rm s}^{\rm (iso)}})^{-1}+(\ten{m_{\rm s}^{\rm (ani)}})^{-1}$ the total effective strained mass. In both cases, the effective mass decreases as the strain increases. Although the trend is perfectly captured by the CB approximation, the numerical agreement is  only moderate. The main problem is that the CB approximation predicts an anisotropic behavior for the mass, since the red and blue solid lines are different, while in the 8B model the mass is isotropic. The reason behind this discrepancy comes from the fact that, in the 8B model, the LHB and HHB are allowed to be mixed, so that, in a way, the effective splitting between them is averaged out. However, in the CB approximation they are treated as independent, decoupled bands.

Figure~\ref{Fig1}(d) shows the effective $g$-factor; $g_{xx}$ versus  $e_{zz}$ in blue, and $g_{zz}$ versus $e_{xx}$ in red. We note that, since the effective $g$-factor comes from an (atomic) orbital effect, the effect of strain in one direction is reflected in a change of the $g$-factor in the perpendicular one. In both directions, the $g$-factor seems to converge to the free electron's $g$-factor for negative strain, i.e., $g_e=2$, since the gap increases and the CB and VBs are weakly hybridized. In contrast, for positive strain the $g$-factor grows quite dramatically. For large positive strains, the CB approximation fails as the $g$-factor diverges due to the closing of the SM gap, a phenomenon that cannot be accounted by simply using L\"owdin perturbation theory.

Finally, in Fig.~\ref{Fig1}(f) we show the SO coupling $\alpha$ normalized to the constant electric field $q_e\partial\phi$. The SO interaction also has an orbital origin, so we plot $\alpha_z$ versus $e_{xx}$ in red and $\alpha_x$ versus $e_{zz}$ in blue. The SO coupling follows a similar trend to the $g$-factor, decreasing with negative strain, where the CB approximation captures well the 8B model calculations, and increasing with positive strain, where the CB approximation fails to capture the 8B model results for large positive values.

\subsection{Biaxial deviatoric deformation}
\label{Sec2-2}
\begin{figure}
    \centering
    \includegraphics[width=1\columnwidth]{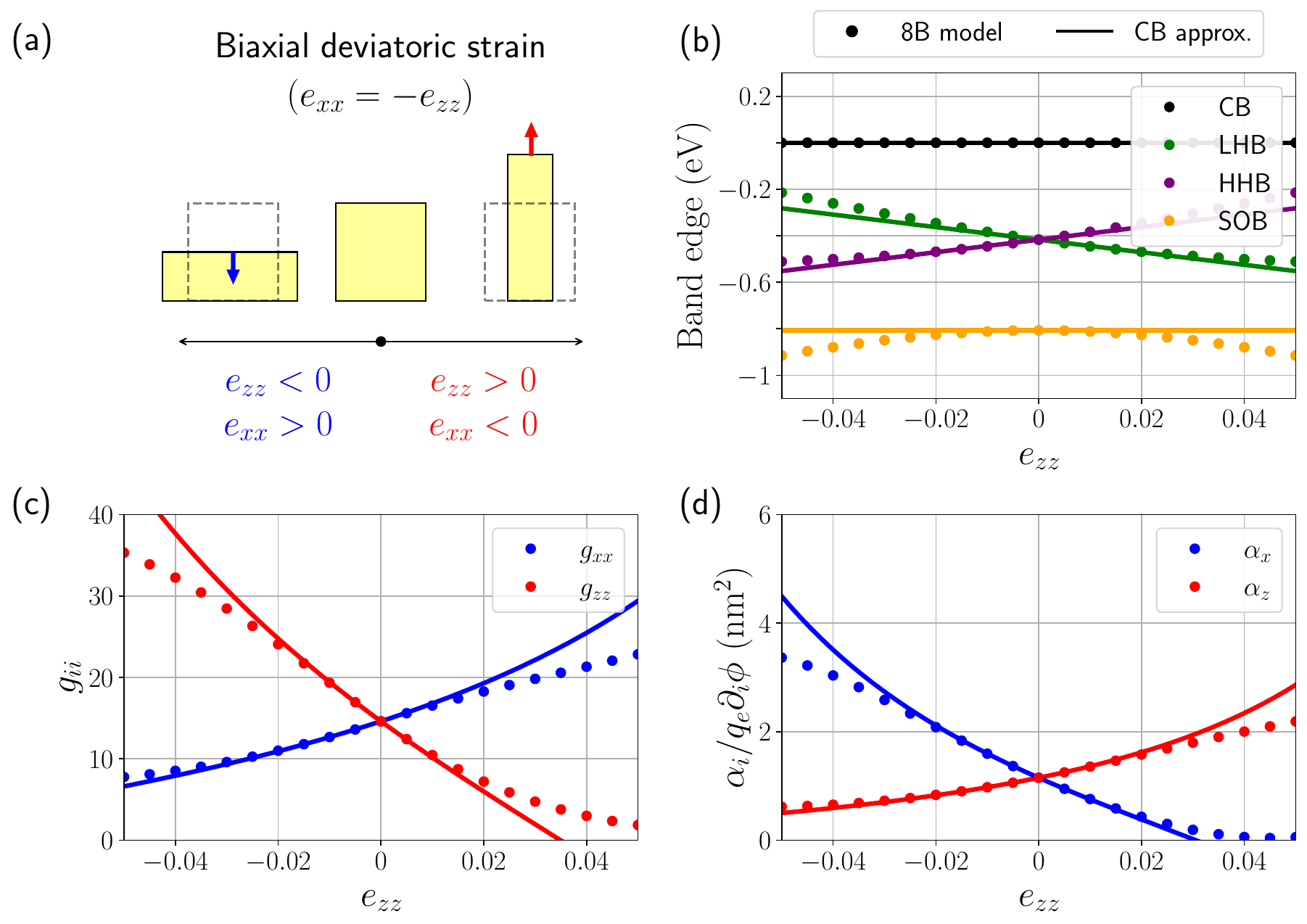}
    \caption{\textbf{Biaxial deviatoric strain.} Sketch of a deviatoric deformation, which is a hydrostatic deformation along two axes that preserves the volume of the unit cell. (b-d) Same as Fig.~\ref{Fig1}(c,e,f), respectively, but for a deviatoric strain along the $z$ direction. The effective mass for this kind of deformation (not shown here) exhibits a similar trend as observed in Fig.~\ref{Fig1}(d) since it does not have an orbital origin.
    %(b) Band edge of the CB (black), LHB (green), HHB (purple) and SOB (orange) versus $e_{zz}$, keeping $e_{xx}=-e_{zz}$. In this, and the following plots, the dots represent the results of the full-diagonalization of the 8-band model Hamiltonian, while the lines represent the results provided by the CB approximation. (c) Effective $g$-factor vs strain. In red we show $g_{zz}$ and in blue $g_{xx}$. (d) Spin-orbit coupling $\alpha$, in units of electric field $q_e\partial\phi$ times nm$^2$, as a function of strain. In blue we show $\alpha_z$ and in red $\alpha_x$. Same parameters as in Fig.~\ref{Fig1}, i.e., for InAs.
    }
    \label{Fig2}
\end{figure}

We now consider a biaxial hydrostatic deformation that preserves the volume of the unit cell. In low-dimensional systems, this kind of deformation is more common than a uniaxial deformation. Restricting it to the $x-z$ plane and applying a deformation in the $z$ direction, the strain tensor is
\begin{eqnarray}
    \ten{e}=\begin{pmatrix}
-e_{zz} & 0 & 0\\
0 & 0 & 0 \\
0 & 0 & e_{zz}
\end{pmatrix},
\end{eqnarray}
so that $\mathrm{tr}\left\{\ten{e}\right\}=0$. Figure~\ref{Fig2}(a) shows a sketch of this kind of deformation, typically called deviatoric in solid-state physics.

As shown in Fig.~\ref{Fig2}(b), this deformation keeps the CB and SOB edges almost unchanged, as well as the mean distance between the LHB and HHB edges, although they split with strain. Again, their relative energy position depends on the sign of the strain. This has a strong impact on the parameters of the system. In Fig.~\ref{Fig2}(c) we show the effective $g$-factor along the two directions affected by the strain. Interestingly, one increases while the other decreases. A similar behavior is observed for the SO coupling, shown in Fig.~\ref{Fig2}(d). We find that for large values of strain, both the $g$-factor and SO coupling are suppressed in one direction (the one compressed) while enhanced in the perpendicular one (the one stretched). In general, the CB approximation turns out to be pretty accurate compared to the full 8B model calculations, except for large strains, i.e., $\left|e_{zz}\right|\gtrsim 0.4$, in which the agreement becomes poorer.

\subsection{Pure shear deformation}
\label{Sec2-3}
\begin{figure}
    \centering
    \includegraphics[width=1\columnwidth]{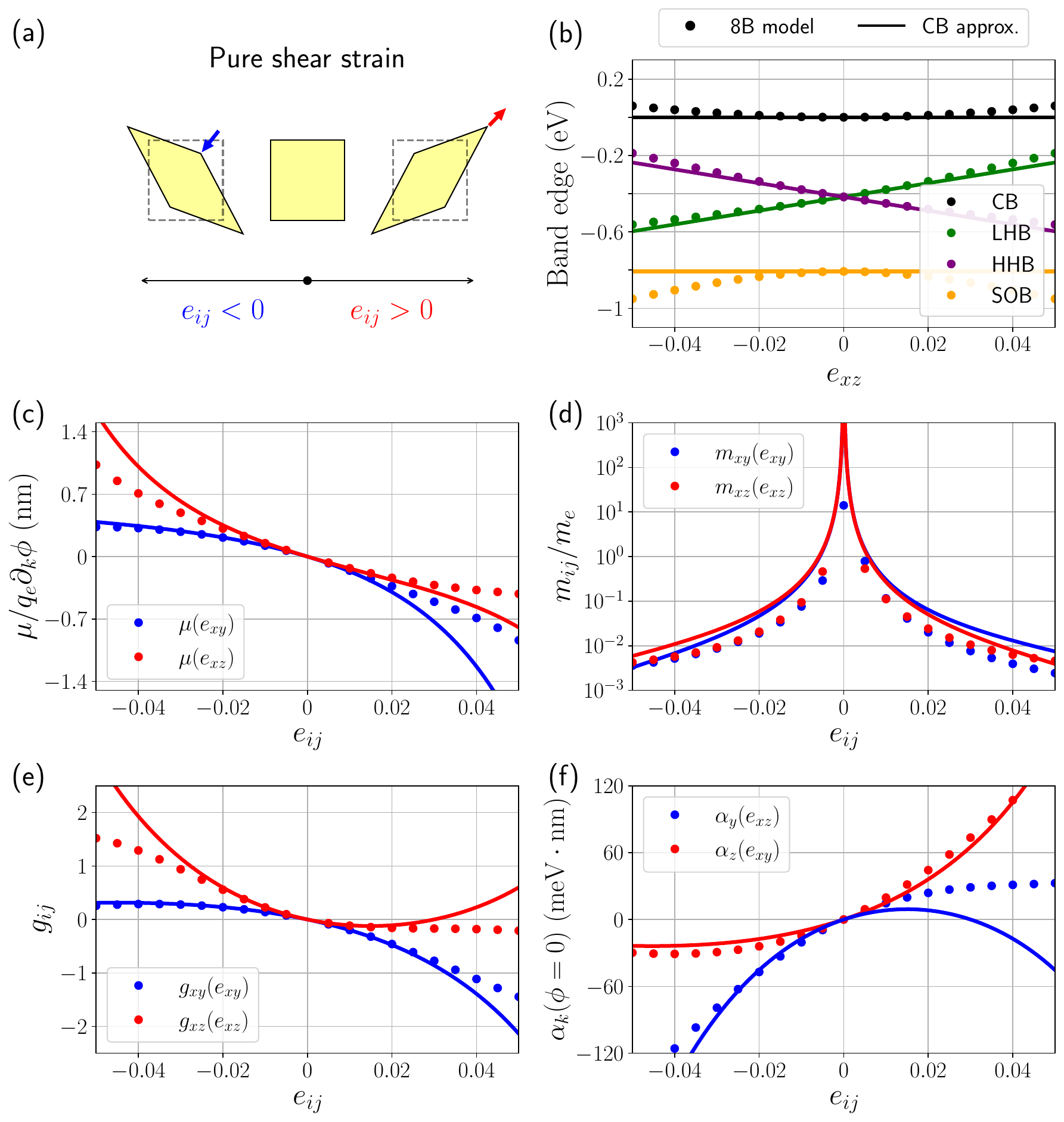}
    \caption{ \textbf{Pure shear strain.} (a) Sketch of a pure shear deformation, which affects only the inner angles of the unit cell, changing its shape but not the size of its borders or the volume. This deformation only involves the non-diagonal terms of the strain tensor, $e_{ij}$ for $i\neq j$. (b,d-f) Same as Fig.~\ref{Fig1}(b,d-f), respectively, but for shear pure strain along the $x-y$ and $x-z$ directions. Notice that in (f), the SO coupling is evaluated at zero electric field. (c) Chemical potential $\mu$, normalized to $q_e\partial\phi$, as a function of strain $e_{xy}$ in blue and $e_{xz}$ in red.}
    %(b) Band edge of the conduction band (black), light-hole valence band (green), heavy-hole valence band (purple) and split-off valence band (orange) vs $e_{xz}$. In this, and the following plots, the dots represent the results of the full-diagonalization of the 8-band model Hamiltonian, while the lines represent the results provided by the CB approximation. (c) Chemical potential $\mu$, in units of electric field $q_e\partial\phi$ times nm, as a function of strain, $e_{xy}$ in blue and $e_{xz}$ in red. (d) Effective mass as a function of strain. In red we show $m_{xy}$ vs $e_{xy}$, while in blue $m_{xz}$ vs $e_{xz}$. (e) Effective $g$-factor vs strain. In red we show $g_{xy}$ vs $e_{xy}$, and in blue $g_{xz}$ vs $e_{xz}$. (f) Spin-orbit coupling $\alpha$, at zero electric field, as a function of strain. In blue we show $\alpha_z$ as a function of $e_{xy}$, and in red $\alpha_y$ as a function of $e_{xz}$. Same parameters as in Fig.~\ref{Fig1} (InAs).}
    \label{Fig3}
\end{figure}

Lastly, we consider the case of a pure shear deformation that only involves the non-diagonal elements of the strain tensor. This kind of deformation only affects the inner angles of the unit cell, altering its shape without changing its volume nor the size of its borders (up to $e^2$ order). A sketch of this deformation is depicted on Fig.~\ref{Fig3}(a).  The strain tensor (up to quadratic terms) reads
\begin{eqnarray}
    \ten{e}\simeq\begin{pmatrix}
0 & e_{xy} & e_{xz}\\
e_{xy} & 0 & e_{yz} \\
e_{xz} & e_{yz} & 0
\end{pmatrix}.
\end{eqnarray}
For brevity of the discussion, we will take $e_{yz}=0$ and analyze the effect of changing $e_{xy}$ and $e_{xz}$ (the change with $e_{yz}$ should be equivalent to $e_{xz}$ due to crystal symmetry).  

The band edges as a function $e_{xz}$ are shown in Fig.~\ref{Fig3}(b). Similarly to the biaxial deviatoric case, the CB and SOB tend to remain constant while a splitting is created between the HHB and LHB with increasing $e_{xz}$. In these cases, nonetheless, there is a small change in the SOB and CB edges that the CB approximation cannot capture. The CB approximation, however, does account for a change in the chemical potential due to $e_{xz}$, but only when an electric field is considered, similarly to a Darwin correction~\cite{Winkler2003}. In Fig.~\ref{Fig3}(c) we show the change of the chemical potential in units of an electric field parallel to the shear component $q_e\partial\phi$. The sign of this chemical potential depends on both the sign of the strain and the electric field. For small and moderate strain, the CB approximation appears to successfully predict the behavior.

We show the strained effective mass in Fig.~\ref{Fig3}(d). The shear strain affects mainly the non-diagonal terms of the effective mass, $m_{xy}$ in blue and $m_{xz}$ in red. Notice that for the non-diagonal terms, an infinite effective mass means no kinetic energy term coming from $k_x\frac{\hbar^2}{m_{xy}}k_y$ terms. We observe that for large strains the non-diagonal effective mass is of the same order of magnitude than the diagonal terms. Also, the CB approximation seems to account properly for the order of magnitude. We remark that a non-diagonal effective mass introduces linear terms in low-dimensional systems, contributing to a Dirac-like behavior for low-densities.

Finally, in Fig.~\ref{Fig3}(e,f) we show the effective $g$-factor and SO coupling (at zero electric field), respectively. Both follow a similar trend, albeit with opposite signs with respect to $e_{ij}$. The CB approximation provides only a  moderate agreement with the full 8B model calculations in both cases, and it is specially good for small strain, i.e., $\left|e_{ij}\right|\lesssim 0.3$.

\section{Low-dimensional nanostructures}
\label{Sec3}

Strain effects play a significant role in low-dimensional systems. 2D and 1D transport channels are typically realized in 2DEGs or in NWs. These structures often consist of multiple SM layers grown on top of one another to confine the active transport region to a narrow, low-dimensional area. Since different SM materials generally have different lattice constants, even carefully selected combinations designed to minimize mismatch inevitably introduce strain into the active layer~\cite{Micron:Goodhew99}. Therefore, we can make use of the effective Hamiltonian that we have derived to address this phenomenon in a simple and reliable way. In this section, we apply our model to study representative nanostructures in which strain effects are expected to play a crucial role.

But before discussing the details of each of them, let us remark that in low-dimensional systems the electronic spectrum consists of multiple transverse subbands, each of which may experience distinct SO couplings and $g$-factors. In practice, however, experiments usually probe only an effective or averaged response of these parameters, like in magneto-conductance experiments. For this reason, rather than presenting the SO coupling and $g$-factor of each individual subband, we report here their expectation values, computed as
\begin{eqnarray}
    \left<\vec{\alpha}_{\rm s}\right> = \frac{\sum_j \left<\Psi^{(j)}(\vec{r})\right|\vec{\alpha}_{\rm s}^{(j)}(\vec{r})\left|\Psi^{(j)}(\vec{r})\right>n^{(j)}}{\sum_j n^{(j)}},
\end{eqnarray}
where $\Psi^{(j)}$ denotes the eigenfunction for each subband $j$ and $n^{(j)}$ its occupation for a given temperature. The expectation value for the $g$-factor is computed in an analogous manner. This averaging procedure has been shown~\cite{PRR:Escribano20} to yield a good quantitative agreement with experimental measurements.

\begin{figure}
    \centering
    \includegraphics[width=1\columnwidth]{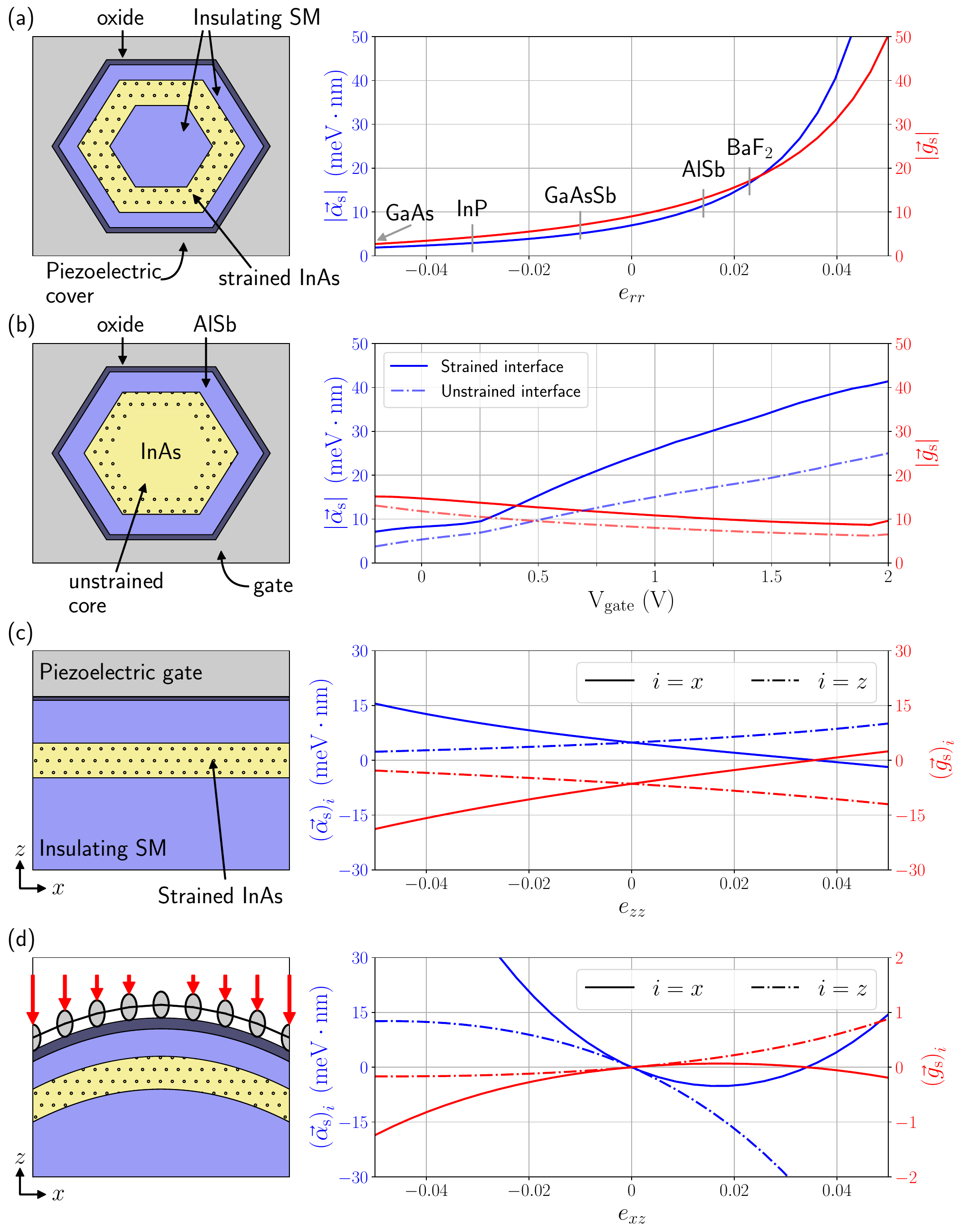}
    \caption{ \textbf{Strain effects in low-dimensional devices.} (a) On the left, sketch of a core-double shell NW with hexagonal cross section. The middle shell, made of InAs, is homogeneously strained due to the inner core and outer shell, made of an insulating material. The NW can be wrapped by an additional metallic layer acting as a gate, or by a piezoelectric material to tune the strain. On the right, modulus of the simulated NW SO coupling (blue) and $g$-factor (red) vs the radial strain $e_{rr}$ induced on the InAs shell. This strain can be created through the piezoelectric gate or appear naturally due to the lattice mismatch with the core and outer shell. Different materials for the insulating SM, which induce different radial strains, are shown. (b) Same as (a) but for a nanostructure without the insulating core. In this case, the strain only affects a narrow region close to the InAs/AlSb interface. The outer gate dopes the NW and also pushes the wavefunction towards the interface, increasing the strain effects. The SO coupling and the $g$-factor are shown against the voltage applied to the outer gate. Results for strained and unstrained interfaces are shown. (c) Same than (a) but for a planar SM heterostructure that electrostatically defines a 2DEG. A piezoelectric gate can be used on top to regulate the strain in the $z$ direction, which induces a deviatoric strain in the $x$ direction as well. Solid lines correspond to the $x$ component of $\vec{\alpha}_{\rm s}$ (blue) and $\vec{g}_{\rm s}$ (red), while dashed lines correspond to their $z$ component. (d) Same structure as (c) but now a lattice of piezolectric gates (in gray) bends the planar heterostructure inducing a shear deformation in the $x$-$z$ plane. A list with all the parameters used for these simulations can be found in Appendix~\ref{AP:parameters}.}
    \label{Fig4}
\end{figure}

We begin by analyzing a core-double shell NW, as depicted in Fig.~\ref{Fig4}(a) (left panel). In this design, the inner core and outer shell (shown in blue) are typically composed of insulating SMs, which serve to confine the electronic states within the inner shell (yellow), i.e., the active layer. The magnitude of the strain experienced by the active layer depends on the lattice mismatch between the different materials used. To control the chemical potential within the device, the outermost layer is often coated with a metallic gate. Here, we propose an alternative (yet closely related) architecture in which this outermost layer is made of a piezoelectric material. By driving a current through it, the piezoelectric layer can expand or contract, thereby inducing controllable strain throughout the NW.

In the right panel of Fig.~\ref{Fig4}(a), we illustrate how both the SO coupling modulus (of the expectation value) and the effective $g$-factor vary with the applied strain in the active layer, assuming it is made of InAs and subjected to hydrostatic deformation\footnote{Although this is a good approximation, the actual strain distribution is more complex near the corners of the hexagonal cross section, as observed in Ref.~\onlinecite{NatCom:Balaghi19}.}. Our results indicate that compressive strain significantly reduces both the SO coupling and the $g$-factor, while tensile strain enhances them, potentially by a factor of two or three. Such strain can be tuned continuously via the piezoelectric shell, or discretely by selecting appropriate insulating materials. In Fig.~\ref{Fig4}(a), we indicate the resulting strain values for several common choices of insulating materials (GaAs, InP, etc.) that have already been successfully grown on InAs. Notably, ternary compounds like GaAs$_x$Sb$_{1-x}$ offer a wider flexibility~\cite{SSC:Andersson87}, enabling strain optimization by fine-tuning their lattice constant through the partial composition $x$.

A simpler nanodevice architecture consists of a single core–shell NW, as shown in the left panel of Fig.~\ref{Fig4}(b), which is used for a wide range of applications~\cite{NanoLet:Skold05, JAP:Gronqvist09, NanoLet:Hetzl16, NatCom:Balaghi19, NatCom:Balaghi21}. In this configuration, the outer metallic layer functions as a gate to tune the chemical potential of the system. It can also be used to leverage the strain-modified parameters, as we show below. The inner core region is typically unstrained, depending on the NW radius. Due to the large surface-to-volume ratio in NWs, strain tends to relax rapidly, typically within 5 to 10~nm, so only a narrow region near the core–shell interface experiences significant strain. For this reason, and given that our Hamiltonian is valid only under a uniform strain distribution, we approximate the strain profile as a step function that affects only a thin layer near the interface. Consistent with the material choices used in Fig.~\ref{Fig4}(a), we consider a core of InAs and an outer shell of AlSb.

The right panel of Fig.~\ref{Fig4}(b) shows how the SO coupling modulus and the effective $g$-factor vary as a function of the gate potential $V_{\rm gate}$. Solid lines correspond to simulations that include the strain profile described above, while dashed lines represent results obtained without considering any strain. Interestingly, both parameters exhibit opposite trends as a function of $V_{\rm gate}$. In particular, the SO coupling increases significantly with increasing gate voltage, an effect that is strongly amplified by the presence of strain. Moreover, this enhancement is notably nonlinear. This nonlinearity arises because the gate potential influences not only the electric field and carrier density within the NW, but also the spatial distribution of the electronic wavefunction across its cross section. A positive gate potential pushes the wavefunction closer to the core–shell interface, where strain is localized, thereby amplifying its effect on the electronic properties. Thus, this observation could serve as indirect evidence for the presence of a strained interface in core–shell NWs. Finally, we emphasize that selecting an appropriate insulating shell material can double the SO coupling strength. This tunability may have important implications for spintronics~\cite{RMP:Zutik04, NatPhys:Awschalom07} and for the realization of topological superconductivity~\cite{rnc:Aguado17, AdvMat:Kim23}, where strong SO coupling is a key requirement.

We now turn to similar geometries in quasi-2D nanostructures. Figure~\ref{Fig4}(c) illustrates a common design for confining a 2DEG: a layered heterostructure in which an active layer, InAs in this case (yellow), is sandwiched between two identical insulating materials (blue). As in the NW geometries, the active layer may experience strain due to lattice mismatch~\cite{JAP:Jones17, SSC:Andersson87}. However, in this planar configuration, the strain can also be externally tuned, for instance by growing a piezoelectric layer on top of the device~\cite{APL:Shayegan03, PRB:Habib07, PRB:Tholen16} or by mechanically deforming the entire nanostructure~\cite{Micromachines:Alekseev20}. In this case, the resulting strain is likely to be deviatoric, as the system is unconstrained along the $x$-direction and the unit cell tends to preserve its volume. 

The right panel of Fig.~\ref{Fig4}(c) shows the two components of the SO coupling (blue) and $g$-factor (red): one parallel to the 2DEG plane ($i = x$, solid lines) and the other perpendicular to it ($i = z$, dashed lines). Under a positive stress applied in the $z$-direction (perpendicular to the 2DEG), or alternatively positive $e_{zz}$, the parallel components of both the SO coupling and the $g$-factor decrease significantly, approaching negligible values. In contrast, the perpendicular components increase, nearly doubling in magnitude. The opposite behavior is observed under compressive (negative) stress, as expected for a deviatoric deformation. This anisotropic strain response can be exploited to implement a spin field-effect transistor (spin-FET)~\cite{APL:Datta90, NatPhys:Awschalom07}, where spin-dependent transport is controlled by externally applied strain.

Finally, we analyze a related geometry in which the planar heterostructure is bent, as illustrated in Fig.~\ref{Fig4}(d). Such bending can be achieved using a superlattice of piezoelectric gates~\cite{AdvMat:Cakiroglu23} that induce spatially varying deformations along the nanostructure. Alternatively, mechanical stress can be applied directly to produce a similar effect~\cite{LSA:Peng20}. In this configuration, the dominant strain component is of the shear type. For simplicity, we consider a pure shear deformation, although in practice the strain may include a hydrostatic component depending on the specific stress profile.

The right panel of Fig.~\ref{Fig4}(d) presents the same quantities shown in Fig.~\ref{Fig4}(c), the SO coupling components and the $g$-factor, but now plotted as a function of the shear strain component $e_{xz}$. As the shear strain increases, both parameters exhibit non-monotonic behavior at large values of $e_{xz}$, likely signaling the breakdown of the effective Hamiltonian approximation, as discussed in Sec.~\ref{Sec1}. Nevertheless, this simulation correctly captures the pronounced dependence of the SO coupling with shear strain even in the absence of an external electric field. This effect arises from the mechanism described in Sec.~\ref{Sec2-3}.

In addition, shear strain induces non-zero off-diagonal components in the $g$-tensor. This suggests that a suitable combination of hydrostatic strain [as in Fig.~\ref{Fig4}(c)] and shear strain [as in Fig.~\ref{Fig4}(d)] could enable the design of nanodevices with an anisotropic $g$-tensor, e.g., with $g_{xx} = 0$ but $g_{xz} \neq 0$. Such a configuration would allow the generation of a Zeeman field along the $z$-direction by applying a magnetic field in the $x$-direction, without inducing a Zeeman splitting in the $x$-direction. Although experimentally achieving this condition would require precise control and fine-tuning of the strain profile, it opens exciting possibilities for engineering topological superconductivity in 2D systems~\cite{PRB:Sau10}.

\section{Conclusions}
\label{Sec4}

We have presented a comprehensive analytical framework to describe the impact of mechanical strain on the conduction band (CB) of zincblende III–V semiconductor (SM) compounds. Starting from the well-established 8-band (8B) k$\cdot$p model including strain, we perform a folding-down procedure of the bands to include the effect of the valence bands (VBs) into the CB. With this, we derive an effective CB Hamiltonian that incorporates strain effects through renormalized parameters, valid under the assumption of uniform strain. We obtain analytical equations for strain-modified quantities such as effective mass, chemical potential, spin–orbit (SO) coupling, and $g$-factor, all of which play crucial roles in the operation of SM nanodevices.

We validate our approach by comparing it to full numerical diagonalizations of the 8B Hamiltonian, finding good agreement for small and moderate strain magnitudes. We have analyzed a variety of deformation types, such as hydrostatic, deviatoric, and pure shear, and we have elucidated their distinct effects on key electronic parameters. Notably, we observe strong strain anisotropy, particularly in the SO interaction and the $g$-tensor, with certain components either enhanced or suppressed depending on the type and direction of the applied strain. Although our study in the main text focuses on InAs, a detailed compilation of effective parameters for the broader set of III–V SM compounds is provided in Appendix~\ref{AP:other}.

Beyond bulk systems, we have applied our model to representative low-dimensional nanostructures, including core–shell nanowires (NWs) and 2DEG heterostructures, showing how strain, either intrinsic due to lattice mismatch or externally induced via piezoelectric gates, can be leveraged to tune device properties. In particular, we demonstrate that both the SO coupling and the $g$-factor can be significantly enhanced or suppressed depending on the strain profile. This may allow to design and develop a spin-FET, where the tunability of the SO interaction is key. In addition, it can help to design the next generation of topological nanodevices, where a strong SO coupling is needed to provide a robust topological protection.

Finally, we emphasize that the core-double shell NW architecture of Fig.~\ref{Fig0}(a) illustrates how our framework can be translated into a realistic straintronic device. By combining intrinsic strain due to lattice mismatch with externally tunable stress provided by a piezoelectric shell, such NWs offer a flexible route to engineer SO coupling and $g$-factors on demand. This level of control not only demonstrates the practical relevance of our effective model, but also highlights a concrete path toward next–generation quantum devices where strain becomes a key design parameter rather than an uncontrolled byproduct.

\begin{acknowledgments}
The authors thank Keiko Takase for the insightful discussions. This work was supported by Grants PID2021-122769NB-I00, PID2021-125343NB-I00, PID2023-150224NB-I00 and PID2024-161665NB-I00 funded by MICIU/AEI/10.13039/501100011033, ``ERDF A way of making Europe'' and ``ESF+'', and through the ``María de Maeztu'' Programme for Units of Excellence in R\&D (CEX2023-001316-M).
\end{acknowledgments}

%\bibliography{biblio}
%%%%%%%%%%%%%%%%%%%%%%%%%%%%%%%%%%%%%%%%%%
%\\

%%%%%%%%%%%%%%%%%%%%%%%%%%%%%%%%%%%%%%%%%%%%%%
\newpage

\appendix
\onecolumngrid

\section{Derivation of the effective strained conduction-band Hamiltonian}
\label{AP:eqs}
Our starting point is the 8-band (8B)
k$\cdot$p model Hamiltonian~\cite{JPCS:Kane57, Winkler2003}, which accurately reproduces the band structure and electronic properties of various semiconductors (SMs)~\cite{JAP:Vurgaftman01}, particularly the III–V compounds considered in this work. The Hamiltonian explicitly includes the four spinful bands closest to the SM band gap [see Fig.~\ref{Fig0}(a)], i.e., the conduction band (CB), light-hole valence band (LHB), heavy-hole valence band (HHB) and split-off valence band (SOB); while higher-energy bands are incorporated as first-order corrections via L\"owdin perturbation theory. Owing to the symmetry of the zincblende unit cell, these four bands can be directly related to the atomic valence orbitals as
\begin{eqnarray}
\left\{ \begin{array}{ll}
\left| \mathrm{CB}_{\uparrow} \right>=\left| S_{\uparrow} \right> , &
\left| \mathrm{CB}_{\downarrow} \right>=\left| S_{\downarrow} \right>, \\
\left| \mathrm{LHB}_{\uparrow} \right>=\frac{i}{\sqrt{6}}\left| (X+iY)_{\downarrow} -2Z_{\uparrow}\right>, &
\left| \mathrm{LHB}_{\downarrow} \right>=\frac{1}{\sqrt{6}}\left| (X-iY)_{\uparrow} +2Z_{\downarrow}\right>, \\
\left| \mathrm{HHB}_{\uparrow} \right>=\frac{1}{\sqrt{2}}\left| (X+iY)_{\uparrow} \right>, &
\left| \mathrm{HHB}_{\downarrow} \right>=\frac{i}{\sqrt{2}}\left| (X-iY)_{\downarrow} \right>, \\
\left| \mathrm{SOB}_{\uparrow} \right>=\frac{1}{\sqrt{3}}\left| (X+iY)_{\downarrow} +Z_{\uparrow}\right>, &
\left| \mathrm{SOB}_{\downarrow} \right>=\frac{i}{\sqrt{3}}\left| -(X-iY)_{\uparrow}+Z_{\downarrow} \right>,
\end{array}
   \right.
\end{eqnarray}
where $S$, $X$, $Y$ and $Z$ denote the type of symmetry ($s$-function, or $p$-function $x$, $y$ or $z$) that the orbital has under the tetrahedral group transformation, and $\left\{\uparrow,\downarrow\right\}$ denotes the spin direction. In this basis, the $z$ direction corresponds to the $(001)$ crystallographic orientation. The 8B Hamiltonian in the $\Psi=(\Psi_{\rm{CB},\uparrow},\Psi_{\rm{CB},\downarrow},\Psi_{\rm{HHB},\uparrow},\Psi_{\rm{LHB},\uparrow},\Psi_{\rm{LHB},\downarrow},\Psi_{\rm{HHB},\downarrow},\Psi_{\rm{SOB},\uparrow},\Psi_{\rm{SOB},\downarrow})$ basis is given by~\cite{JPCS:Kane57, Winkler2003}
\begin{equation}
H_{0}=
  \begin{pmatrix}
    T_{\rm{c}} & 0 & \frac{1}{\sqrt{6}}Pk_+ & 0 & \frac{1}{\sqrt{2}}Pk_- & -\sqrt{\frac{2}{3}}Pk_z & -\frac{1}{\sqrt{3}}Pk_z & \frac{1}{\sqrt{3}}Pk_+  
    \\
    
    0 & T_{\rm{c}} & -\sqrt{\frac{2}{3}}Pk_z & -\frac{1}{\sqrt{2}}Pk_+ & 0 &  -\frac{1}{\sqrt{6}}Pk_- & \frac{1}{\sqrt{3}}Pk_- & \frac{1}{\sqrt{3}}Pk_z     
    \\
    
    \frac{1}{\sqrt{6}}Pk_- & -\sqrt{\frac{2}{3}}Pk_z & T_{\rm{lh}} & -\Omega_2^\dagger & \Omega_1 & 0 & \sqrt{\frac{3}{2}}\Omega_2 & -\sqrt{2}\Omega_3 
    \\
    
    0 & -\frac{1}{\sqrt{2}}Pk_- & -\Omega_2 & T_{\rm{hh}} & 0 & \Omega_1 & -\sqrt{2}\Omega_1 & \frac{1}{\sqrt{2}}\Omega_2 
    \\
    
    \frac{1}{\sqrt{2}}Pk_+ & 0 & \Omega_1^\dagger & 0 & T_{\rm{hh}} & \Omega_2^\dagger & \frac{1}{\sqrt{2}}\Omega_2^\dagger & \sqrt{2}\Omega_1^\dagger  
    \\

    -\sqrt{\frac{2}{3}}Pk_z & -\frac{1}{\sqrt{6}}Pk_+ & 0 & \Omega_1^\dagger & \Omega_2 & T_{\rm{lh}} & \sqrt{2}\Omega_3 & \sqrt{\frac{3}{2}}\Omega_2^\dagger  
    \\

    -\frac{1}{\sqrt{3}}Pk_z & \frac{1}{\sqrt{3}}Pk_+ & \sqrt{\frac{3}{2}}\Omega_2^\dagger & -\sqrt{2}\Omega_1^\dagger & \frac{1}{\sqrt{2}}\Omega_2 & \sqrt{2}\Omega_3 & T_{\rm{soff}} & 0  
    \\
    
    \frac{1}{\sqrt{3}}Pk_- & \frac{1}{\sqrt{3}}Pk_z & -\sqrt{2}\Omega_3 & \frac{1}{\sqrt{2}}\Omega_2^\dagger & \sqrt{2}\Omega_1 & \sqrt{\frac{2}{3}}\Omega_2 & 0 & T_{\rm{soff}}  
    \\
    
  \end{pmatrix},\label{H_8B_0}
\end{equation} 
where the diagonal terms are
\begin{eqnarray}
T_{\rm{c}}=E_{\rm F}+\frac{\hbar^2\vec{k}^2}{2m_e}-e\phi(\vec{r}), \\
T_{\rm{lh}}=E_{\rm h}-\frac{\hbar^2}{2m_e}\left[(k_x^2+k_y^2)(\gamma_1-\gamma_2)+k_z^2(\gamma_1+2\gamma_2)\right]-e\phi(\vec{r}),  \\
T_{\rm{hh}}=E_{\rm h}-\frac{\hbar^2}{2m_e}\left[(k_x^2+k_y^2)(\gamma_1+\gamma_2)+k_z^2(\gamma_1-2\gamma_2)\right)-e\phi(\vec{r}),  \\
T_{\rm{soff}}=E_{\rm{soff}}-\gamma_1\frac{\hbar^2\vec{k}^2}{2m_e}-e\phi(\vec{r}),
\end{eqnarray}
and the off-diagonal ones are
\begin{eqnarray}
\Omega_1=-\sqrt{3}\frac{\hbar^2}{2m_e}\left[\gamma_2(k_x^2-k_y^2)-2i\gamma_3k_xk_y\right],\\
\Omega_2=-\sqrt{3}\frac{\hbar^2}{m_e}\gamma_3k_zk_-,\\
\Omega_3=\frac{\hbar^2}{2m_e}\gamma_2\left(k_x^2+k_y^2-2k_z^2\right).
\end{eqnarray}
Here, $E_{\rm F}$, $E_{\rm h}=E_F-\Delta_{\rm g}$ and $E_{\rm{soff}}=E_{\rm F}-\Delta_{\rm g}-\Delta_{\rm{soff}}$ are the (unstrained) band edges of the CB, HHB and SOB, respectively; $\Delta_{\rm g}$ and $\Delta_{\rm{soff}}$ are the gaps between the CB/HHB and the HHB/SOB at the $\Gamma$ point; $\phi(\vec{r})$ is the slow-varying electrostatic potential; $\vec{k}$ the envelope momentum, with $k_\pm\equiv k_x\pm ik_y$; and $\left\lbrace\gamma_i\right\rbrace$ and $P$ are Kane parameters. In this work, we choose the CB edge as the reference energy, $E_{\rm F}=0$. The Hamiltonian elements whose functional forms have been substituted by phenomenological parameters are
\begin{eqnarray}
P\equiv-\frac{i\hbar}{m_e}\left<S\right|p_x\left|X\right>=-\frac{i\hbar}{m_e}\left<S\right|p_y\left|Y\right>=-\frac{i\hbar}{m_e}\left<S\right|p_z\left|Z\right>,  \\
\Delta_{\rm{soff}}\equiv\frac{3\hbar i}{4m_e^2c^2}\left<X\right|\frac{\partial \phi_{\rm c}}{\partial x}p_y-\frac{\partial \phi_{\rm c}}{\partial y}p_x\left|Y\right>=
\frac{3\hbar i}{4m_e^2c^2}\left<Y\right|\frac{\partial \phi_{\rm c}}{\partial y}p_z-\frac{\partial \phi_{\rm c}}{\partial z}p_y\left|Z\right>= \frac{3\hbar i}{4m_e^2c^2}\left<Z\right|\frac{\partial \phi_{\rm c}}{\partial z}p_x-\frac{\partial \phi_{\rm c}}{\partial x}p_z\left|X\right>,
\end{eqnarray}
with $\phi_{\rm c}$ the electrostatic potential created by the crystal lattice and $\vec{p}$ the crystal momentum. Notice that we are using first-order L\"owdin perturbation theory here. Second-order terms only add corrections to account for crystal asymmetries, which are very small in zincblende crystals. We are also dropping SO interaction terms coming from $k$, which are several orders of magnitude smaller than $p$.

Strain effects can be taken into account perturbatively in this 8B Hamiltonian following the derivation of Bahder~\cite{PRB:Bahder90}. The effect of the strain is to deform the original unit cell of the crystal lattice, described by the cell constants $\vec{c}=(c_x,c_y,c_z)$, to a different \textit{strained} unit cell with lattice constants
\begin{equation}
    \vec{c}_{\rm s}=(1+\ten{e})\vec{c},
\end{equation}
where
\begin{equation}
    \ten{e}=\begin{pmatrix}
e_{xx} & e_{xy} & e_{xz}\\
e_{yx} & e_{yy} & e_{yz} \\
e_{zx} & e_{zy} & e_{zz}
\end{pmatrix} \; \rightarrow \;     (\ten{e})_{ij}=e_{ij}\equiv\frac{(\vec{c}_{\rm s})_i-(\vec{c})_{i}}{(\vec{c})_{j}}.
\end{equation}
Each element of this tensor $e_{ij}$ provides the relative change of the unit cell along the $i$ direction when a deformation is applied to the $j$ one. The diagonal elements are called \emph{normal} strains (and the trace $\mathrm{tr}\left\{\ten{e}\right\}$ provides the so-called hydrostatic strain), while the non-diagonal terms are called \emph{shear} strains.

Strain modifies the unperturbed crystal potential $\phi_{\rm c}(\vec{r})$ into one that now possesses the new periodicity of the strained unit cell, $\phi_{\mathrm{c,s}}(\vec{r})$. If the strain $\ten{e}$ is small, then one can expand perturbatively (to first order) the crystal electrostatic potential
\begin{equation}
    \phi_{\mathrm{c,s}}(\vec{r})\simeq \phi_{\rm c}(\vec{r})+\sum_{i,j}  e_{ij} \left(\frac{\partial \phi_{\mathrm{c,s}}(\vec{r})}{\partial e_{ij}}\right)_{e_{ij}=0}.
\label{Eq:potential_strain}
\end{equation}
This allows us to introduce the strain also as an additional (perturbative) term in the 8B k$\cdot$p Hamiltonian. This is the so-called Bir-Pikus approach~\cite{Bir1974}, and provides the correction to the Hamiltonian
\begin{equation}
H_{\rm{s}}=
  \begin{pmatrix}
    0 & 0 & t^*-v^* & 0 & -\sqrt{3}(t+v) & \sqrt{2}(w+u) & w+u & \sqrt{2}(t^*-v^*)  
    \\
    
    0 & 0 & \sqrt{2}(w+u) & -\sqrt{3}(t^*-v^*) & 0 &  t+v & -\sqrt{2}(t+v) & w^*-u^*     
    \\
    
    t-v & \sqrt{2}(w^*+u^*) & t_{\rm lh} & -\omega_2^* & \omega_1 & 0 & \sqrt{\frac{3}{2}}\omega_2 & -\sqrt{2}\omega_3  
    \\
    
    0 & -\sqrt{3}(t-v) & -\omega_2 & t_{\rm hh} & 0 & \omega_1 & -\sqrt{2}\omega_1 & \frac{1}{\sqrt{2}}\omega_2 
    \\
    
    -\sqrt{3}(t^*+v^*) & 0 & \omega_1^* & 0 & t_{\rm hh} & \omega_2^* & \frac{1}{\sqrt{2}}\omega_2^* & \sqrt{2}\omega_1^*  
    \\

    \sqrt{2}(w^*+u^*)  & t^*+v^* & 0 & \omega_1^* & \omega_2 & t_{\rm lh} & \sqrt{2}\omega_3 &  \sqrt{\frac{3}{2}}\omega_2^* 
    \\

    w^*+u^*  & -\sqrt{2}(t^*+v^*) & \sqrt{\frac{3}{2}}\omega_2^* & -\sqrt{2}\omega_1^* & \frac{1}{\sqrt{2}}\omega_2 & \sqrt{2}\omega_3 & t_{\rm soff} & 0  
    \\
    
    \sqrt{2}(t-v) & w-u & -\sqrt{2}\omega_3 & \frac{1}{\sqrt{2}}\omega_2^* & \sqrt{2}\omega_1 & \sqrt{\frac{3}{2}}\omega_2 & 0 & t_{\rm soff}
    \\
    
  \end{pmatrix}. \label{H_8B_s}
\end{equation} 
By direct inspection of Eq.~\eqref{Eq:potential_strain}, this perturbation to the Hamiltonian should have the same functional form as the unperturbed one but with different prefactors, which should moreover be proportional to the strain $e_{ij}$. Particularly, they are found to be
\begin{eqnarray}
    t = \frac{1}{\sqrt{6}}d_{\rm cv}(e_{xz}+ie_{yz}), \\
    w = \frac{i}{\sqrt{3}}d_{\rm cv}e_{xy}, \\
    u = \frac{1}{\sqrt{3}}P\sum_j e_{zj}k_j, \\
    v = \frac{1}{\sqrt{6}}P\sum_j (e_{xj}-ie_{yj})k_j;
\end{eqnarray}
\begin{eqnarray}
    t_{\rm lh} = -(e_{xx}+e_{yy})\left(a+\frac{b}{2}\right)-e_{zz}(a-b), \\
    t_{\rm hh} =  -(e_{xx}+e_{yy})\left(a-\frac{b}{2}\right)-e_{zz}(a+b), \\
    t_{\rm soff} = -a(e_{xx}+e_{yy}+e_{zz});
\end{eqnarray}
\begin{eqnarray}
    \omega_1 = \frac{\sqrt{3}}{2}b(e_{xx}-e_{yy})-id_{\rm v}e_{xy}, \\
    \omega_2 = -d_{\rm v}(e_{xz}-ie_{yz}), \\
    \omega_3 = b\left[e_{zz}-\frac{1}{2}(e_{xx}+e_{yy})\right];
\end{eqnarray}
with $a$, $b$, $d_{\rm cv}$ and $d_{\rm v}$ the additional (phenomenological) Kane parameters due to strain. In the literature, $a$ is typically referred to as the deformation potential constant and, in our sign convention $a>0$, so that a positive strain represents a stretching of the unit cell.

The Hamiltonian describing the 8Bs of the SM under strain is therefore given by $H_0+H_{\rm s}$ [Eqs.~\eqref{H_8B_0} and~\eqref{H_8B_s}]. This formulation includes the 8Bs closest to the Fermi level. However, our goal is to obtain a simpler and more manageable equation involving only the CB. To achieve this, we further reduce the model by performing again a L\"owdin partitioning (band folding)
\begin{equation}
   H_0+H_{\rm s} =\begin{pmatrix}
H_{\rm c} & H_{\rm cv} \\
H_{\rm cv}^\dagger & H_{\rm v}
\end{pmatrix} \; \rightarrow \; H_{\rm CB, s} = H_{\rm c} + H_{\rm cv}G_{\rm v} H_{\rm cv}^\dagger,
\end{equation}
where $G_{\rm v}=\left(E-H_{\rm v}\right)^{-1}$ denotes the Green’s function of the VB. Here, $H_{\rm c}$ and $H_{\rm v}$ are the block Hamiltonians corresponding to the conduction and valence sectors of $H_0+H_{\rm s}$, while $H_{\rm cv}$ represents their coupling. Notice that matrix inversion in the Green’s function does not allow for an analytical expression in general. We thus perform a Dyson expansion assuming that its diagonal elements, i.e., the energy gaps, dominate over the off-diagonal ones, i.e., the coupling among the VBs [see Ref.~\onlinecite{PRR:Escribano20} for a detailed derivation without strain]. At zeroth-order the Green's function becomes diagonal and we can find an analytical expression for the CB, which, after regrouping, we write as
\begin{eqnarray}
H_{\rm CB,s} & \simeq & H_{\rm CB,s}^{\rm (iso)} + H_{\rm CB,s}^{\rm (ani)}, 
\end{eqnarray}
\begin{eqnarray}
H_{\rm CB,s}^{\rm (iso)} & = &   \left(\vec{k}\cdot\frac{\hbar^2}{2\ten{m_{\rm s}^{\rm (iso)}}}\cdot\vec{k}^T+\mu_{\rm s}^{\rm (iso)} (\vec{r})-e\phi(\vec{r})\right)\sigma_0 \nonumber \\ 
&+&\frac{1}{2}\left[\vec{\alpha}_{\rm s}^{\rm (iso)}(\vec{r})\times \vec{k}-\vec{k}\times\vec{\alpha}_{\rm s}^{\rm (iso)}(\vec{r})\right]\cdot\vec{\sigma}^T + \frac{1}{2}\mu_{\rm B} \vec{B}\cdot \ten{g_{\rm s}^{\rm (iso)}}\cdot \vec{\sigma}^T, \label{Eq_iso} \\
H_{\rm CB,s}^{\rm (ani)} & = &  \left(\vec{k}\cdot\frac{\hbar^2}{2\ten{m_{\rm s}^{\rm (ani)}}(\vec{r})}\cdot\vec{k}^T+\mu_{\rm s}^{\rm (ani)} (\vec{r})\right)\sigma_0 \nonumber \\
&+&\frac{1}{2}\left[\vec{\alpha}_{\rm s}^{\rm (ani)}(\vec{r})\times \vec{k}-\vec{k}\times\vec{\alpha}_{\rm s}^{\rm (ani)}(\vec{r})\right]\cdot(0,0,\sigma_z)^T + \frac{1}{2}\mu_{\rm B} \vec{B}\cdot \ten{g_{\rm s}^{\rm (ani)}}\cdot (0,0,\sigma_z)^T. \label{Eq_ani}
\end{eqnarray}
We decompose the Hamiltonian into two contributions: an isotropic term, $H_{\rm CB,s}^{\rm (iso)}$, and an anisotropic one, $H_{\rm CB,s}^{\rm (ani)}$, which arises due to the asymmetry of the zincblende unit cell along the $z$ direction [the (001) axis]. In these expressions, $\vec{k}$ and $\vec{r}$ denote the momentum and position operators, respectively, while $\vec{\sigma}$ represents the Pauli spin matrices acting on the CB. For the derivation, we have used the non-commutation relation of the momentum operator in the presence of a magnetic field, 
\begin{equation}
    \left[k_{l},k_{m}\right]=i\frac{e}{\hbar}B_{n}\varepsilon_{lmn},
\end{equation} 
with $\varepsilon_{lmn}$ the elements of the Levi-Civita tensor.

The first terms in Eqs.~\eqref{Eq_iso} and~\eqref{Eq_ani} are the kinetic energy terms, with $\ten{m_{\rm s}^{\rm (iso)}}$ and $\ten{m_{\rm s}^{\rm (ani)}}$ the effective isotropic and anisotropic masses, respectively, given by 
\begin{equation}
\begin{aligned}
\frac{1}{\ten{m_{\rm s}^{\rm (iso)}}} &= \frac{1}{m_e}+ \frac{2P^2}{3\hbar^2}\left(\ten{1}-\ten{e}\right)^2\left(\frac{\sfrac{1}{2}}{\Delta_{\rm LHB, s}} + \frac{\sfrac{3}{2}}{\Delta_{\rm HHB, s}}+\frac{1}{\Delta_{\rm SOB, s}} \right), \\
 \left(\frac{1}{\ten{m_{\rm s}^{\rm (ani)}}}\right)_{jk}  & =   \frac{P^2}{\hbar^2}\left(\ten{1}-\ten{e}\right)_{j z}\left(\frac{1}{\Delta_{\rm LHB, s}}- \frac{1}{\Delta_{\rm HHB, s}} \right)\left(\ten{1}-\ten{e}\right)_{zk}.
 \end{aligned}
\end{equation}
The second terms, $\mu_{\rm s}^{\rm (iso)} (\vec{r})-e\phi(\vec{r})$ or $\mu_{\rm s}^{\rm (ani)}(\vec{r})$, are the (slow-varying) electrochemical potentials
\begin{equation}
\begin{aligned}
  \mu_{\rm s}^{\rm (iso)} (\vec{r}) & =-\frac{q_{e}Pd_{\rm cv}}{3}\left(\frac{\sfrac{1}{2}}{\Delta_{\rm LHB, s}^2} + \frac{\sfrac{3}{2}}{\Delta_{\rm HHB, s}^2}+\frac{1}{\Delta_{\rm SOB, s}^2} \right)\vec{\nabla}\phi(\vec{r})\cdot\vec{e}_{\rm s}, \\
 \mu_{\rm s}^{\rm (ani)} (\vec{r}) & = -\frac{q_e Pd_{\rm cv}}{2}\left(\frac{1}{\Delta_{\rm LHB, s}^2}-\frac{1}{\Delta_{\rm HHB, s}^2} \right)\partial_z\phi(\vec{r})\cdot\left(\vec{e}_{\rm s}\right)_z.
 \end{aligned}
\end{equation}
The third terms correspond to the SO interaction, with $\vec{\alpha}_{\rm s}^{\rm (iso)}(\vec{r})$ the isotropic SO coupling and $\vec{\alpha}_{\rm s}^{\rm (ani)}(\vec{r})$ the anisotropic one,
\begin{equation}
\begin{aligned}
 \vec{\alpha}_{\rm s}^{\rm (iso)} (\vec{r}) & =  \frac{q_e P^2}{3}\mathrm{adj}\left\{\ten{1}-\ten{e}\right\}\left(\frac{1}{\Delta_{\rm LHB, s}^2}-\frac{1}{\Delta_{\rm SOB, s}^2} \right)\vec{\nabla}\phi(\vec{r}) -\frac{Pd_{\rm cv}}{3}\left(\frac{1}{\Delta_{\rm LHB, s}}-\frac{1}{\Delta_{\rm SOB, s}} \right)\vec{e}_{\rm s},\\
\vec{\alpha}_{\rm s}^{\rm (ani)}(\vec{r}) &=-\frac{q_e P^2}{2}\mathrm{adj}\left\{\ten{1}-\ten{e}\right\}\left(\frac{1}{\Delta_{\rm LHB, s}^2}-\frac{1}{\Delta_{\rm HHB, s}^2} \right)\vec{\nabla}\phi(\vec{r}) +\frac{Pd_{\rm cv}}{2}\left(\frac{1}{\Delta_{\rm LHB, s}}-\frac{1}{\Delta_{\rm HHB, s}} \right)\vec{e}_{\rm s}.
 \end{aligned}
\end{equation}
Finally, the fourth terms are the Zeeman fields, being $\ten{g_{\rm s}^{\rm (iso)}}$ the effective isotropic $g$-factor and $\ten{g_{\rm s}^{\rm (ani)}}$ the anisotropic modification,
\begin{equation}
\begin{aligned}
 \ten{g_{\rm s}^{\rm (iso)}} &= g_e\ten{1}-\frac{2}{3}\frac{2m_eP^2}{\hbar^2}\mathrm{adj}\left\{\ten{1}-\ten{e}\right\} \left(\frac{1}{\Delta_{\rm LHB, s}}-\frac{1}{\Delta_{\rm SOB, s}} \right), \\
\ten{g_{\rm s}^{\rm (ani)}}&= \frac{2m_eP^2}{\hbar^2}\mathrm{adj}\left\{\ten{1}-\ten{e}\right\}\left(\frac{1}{\Delta_{\rm LHB, s}}- \frac{1}{\Delta_{\rm HHB, s}} \right).
 \end{aligned}
\end{equation}

In all of the above equations, $q_e$, $m_e$ and $g_e$ denote the free electron's charge, mass and $g$-factor, respectively; $P$ and $d_{\rm cv}$ are the Kane parameters of the unstrained conduction-to-valence band coupling and its correction due to strain; and the gaps $\Delta_{\rm LHB, s}$, $\Delta_{\rm HHB, s}$ and $\Delta_{\rm SOB, s}$ correspond to the strained LHB, HHB and SOB gaps, given by
\begin{eqnarray}
\Delta_{\rm LHB, s}=\Delta_{\rm g}-\left(a+\frac{b}{2}\right)\mathrm{tr}\left\{\ten{e}\right\}-\frac{3b}{2}e_{zz}-d_{\rm cv}\norm{\vec{e}_{\rm s}}, \label{Eq:LH_edge}\\
\Delta_{\rm HHB, s}=\Delta_{\rm g}-\left(a-\frac{b}{2}\right)\mathrm{tr}\left\{\ten{e}\right\}+\frac{3b}{2}e_{zz}+d_{\rm cv}\norm{\vec{e}_{\rm s}}, \label{Eq:HH_edge} \\
\Delta_{\rm SOB, s}=\Delta_{\rm g}+\Delta_{\rm soff}-\mathrm{tr}\left\{\ten{e}\right\},
\label{Eq:SOff_edge}
\end{eqnarray}
with $\Delta_{\rm g}$ and $\Delta_{\rm soff}$ the unstrained SM and split-off band gaps. The operation $\mathrm{adj}\left\{\cdot\right\}$ represents the adjugate matrix operation, and $\mathbf{1}$ is the $3\times3$ identity tensor, so that
\begin{equation}
    \mathrm{adj}\left\{{\ten{1}-\ten{e}}\right\}=\begin{pmatrix}
1-e_{yy}-e_{zz} & e_{xy} & e_{xz}\\
e_{xy} & 1-e_{xx}-e_{zz} & e_{yz} \\
e_{xz} & e_{yz} & 1-e_{xx}-e_{yy} 
\end{pmatrix}+\mathcal{O}\left(e_{\alpha\beta}^2\right).
\end{equation}
The vector $\vec{e}_{\rm s}=(e_{yz},e_{zx},e_{xy})$ contains the shear strains.

\section{Numerical methods}
\label{AP:num}

To compute the energy spectrum of low-dimensional materials, we employ finite difference methods for both the 8B Hamiltonian and its effective CB counterpart. The procedure consists of discretizing space into a grid along the confined directions, such that the corresponding momentum operator transforms as $k_j\rightarrow-i\partial_j$. Along directions where the system remains translationally invariant, we retain $k_j$ as a good quantum number. The differential operator $\partial_j$ is then replaced by finite differences (central differences to first order) using standard schemes. Finally, the resulting tight-binding–like Hamiltonian is diagonalized using the numerical routines implemented in \texttt{SciPy}. For this workflow, we employ our in-house package~\cite{MajoranaNanowiresQSP_v1}.

%It is important to note, nonetheless, that applying FDMs to the 8B Hamiltonian is more subtle. First, one must ensure that the Hamiltonian remains Hermitian when implementing the substitution $k_j\rightarrow-i\partial_j$ in the presence of an inhomogeneous potential  $\phi(\vec{r})$. In contrast, the CB approximation as expressed in Eqs.~\eqref{Eq_iso} and ~\eqref{Eq_ani} is written to guarantee Hermiticity. Second, in low-dimensional systems, spurious solutions may arise as an artifact of discretization for the 8B Hamiltonian due to . To suppress these, we employ an inhomogeneous mesh following the procedure described in Ref. X.

For bulk SMs, we assume that momentum is a good quantum number along all three spatial directions and diagonalize the Hamiltonian for each value of $k_j$. Nonetheless, to properly capture the effective Zeeman field and SO coupling in the 8B model, it is necessary to introduce a unit cell that incorporates either the spatially-varying magnetic vector potential (for the Zeeman field) or the gradient of the electrostatic potential (for the SO coupling). To this end, we adopt a minimal unit cell consisting of three sites with periodic boundary conditions. Within this unit cell, we construct the tight-binding Hamiltonian following the procedure outlined above and subsequently impose periodic boundary conditions across the entire system to recover the bulk spectrum. We then diagonalize the resulting Hamiltonian, as explained before, for each $k_j$.

To extract the effective parameters from the 8B Hamiltonian, we fit the spectrum obtained from the 8B calculation as a function of $k_i$ to the corresponding spectrum of the effective CB equation, which for a bulk system along a single direction is given by the expression
\begin{equation}
\left\{ \begin{array}{ll}
    E(e_{ii},k_{i})=\left(\frac{\hbar^2k_i^2}{2m_{ii}(e_{ii})}-\mu(e_{ii})\right)\pm\sqrt{\left(\alpha_j(e_{ii})k_i\right)^2+\left(\frac{1}{2}\mu_{\rm B}g_{jj}(e_{ii})B_j\right)^2} & \leftarrow\mathrm{Hydrostatic}, \\
    E(e_{ij},k_{i},k_j)=\left(\frac{\hbar^2k_ik_j}{2m_{ij}(e_{ij})}-\mu(e_{ij})\right)\pm\sqrt{\left(\alpha_l(e_{ij})k_j\right)^2+\left(\frac{1}{2}\mu_{\rm B}g_{ij}(e_{ij})B_j\right)^2} & \leftarrow\mathrm{Shear}.
 \end{array} \right.
\end{equation}
The direction $j$ denotes the axis along which an inhomogeneous electrostatic potential $\phi(r_j)$ is applied or, equivalently, the axis of the magnetic field $B_j$, which is necessarily perpendicular to $k_i$. To extract the effective parameters for the hydrostatic case, we first set both the electrostatic potential and magnetic field to zero and determine the effective mass $m_{ii}$. Next, we apply a weak magnetic field ($B_j=0.1$~T) and extract the $g$-factor $g_{jj}$ from the corresponding Zeeman splitting. Finally, we introduce an electrostatic potential by applying an electric field ($-q_e\partial_j\phi(r_j)=1$~meV/nm) and fit the resulting spectrum solely for the SO coupling while keeping the previously obtained parameters fixed. For a shear deformation, we follow the same procedure, but now the direction of $\alpha_l$ can be any (although perpendicular to $k_j$). This procedure is repeated for each strain value $e_{ii}$ or $e_{ij}$.

To realistically describe low-dimensional systems, we estimate the electrostatic potential profile $\phi(\vec{r})$ by solving the Poisson equation
\begin{equation}
    \vec{\nabla}\cdot\left(\epsilon(\vec{r})\vec{\nabla}\phi(\vec{r})\right)=\rho(\vec{r}), \label{Eq_elecpot}
\end{equation}
within the Thomas–Fermi approximation. In this framework, the charge density $\rho(\vec{r})$ is taken as that of a 3D electron gas and must be computed self-consistently together with $\phi(\vec{r})$. As shown in Ref.~\onlinecite{Escribano:tesis}, this provides a reliable approximation for low-dimensional III–V SM compounds. We compute $\phi(\vec{r})$ from Eq.~\eqref{Eq_elecpot} using finite element methods following the routines described in Ref.~\onlinecite{Escribano:tesis} and implemented in Ref.~\onlinecite{MajoranaNanowiresQSP_v1}. Notice that the environment is taken into account through the spatially varying dielectric permittivity $\epsilon(\vec{r})$ and the boundary conditions imposed at the metallic interfaces.

\section{Effective parameters of III-V semiconductor compounds other than InAs}
\label{AP:other}
In this section, we investigate how  the effective parameters vary with strain for different III–V SM compounds. It is important to note that the CB approximation is not applicable to all members of the III–V family. Compounds with an indirect gap, such as GaP or those containing Al, cannot be treated within this framework. In addition, nitride compounds and InP, which exhibit the largest band gaps in the family, show effective parameters that are largely insensitive to strain and are therefore omitted from our analysis. This leaves InAs, InSb, GaAs, and GaSb, which are also the most widely used III–V SMs. Ternary compounds such as InAsSb or GaAsSb could, in principle, be described within the same formalism by interpolating the band gaps and Kane parameters of their binary constituents~\cite{JAP:Vurgaftman01}. Nonetheless, we expect the approach to break down in InAsSb even under small strain, due to its exceptionally small band gap within the family. Hence, we thus restrict ourselves to the binaries.

\begin{figure}
    \centering
    \includegraphics[width=1\columnwidth]{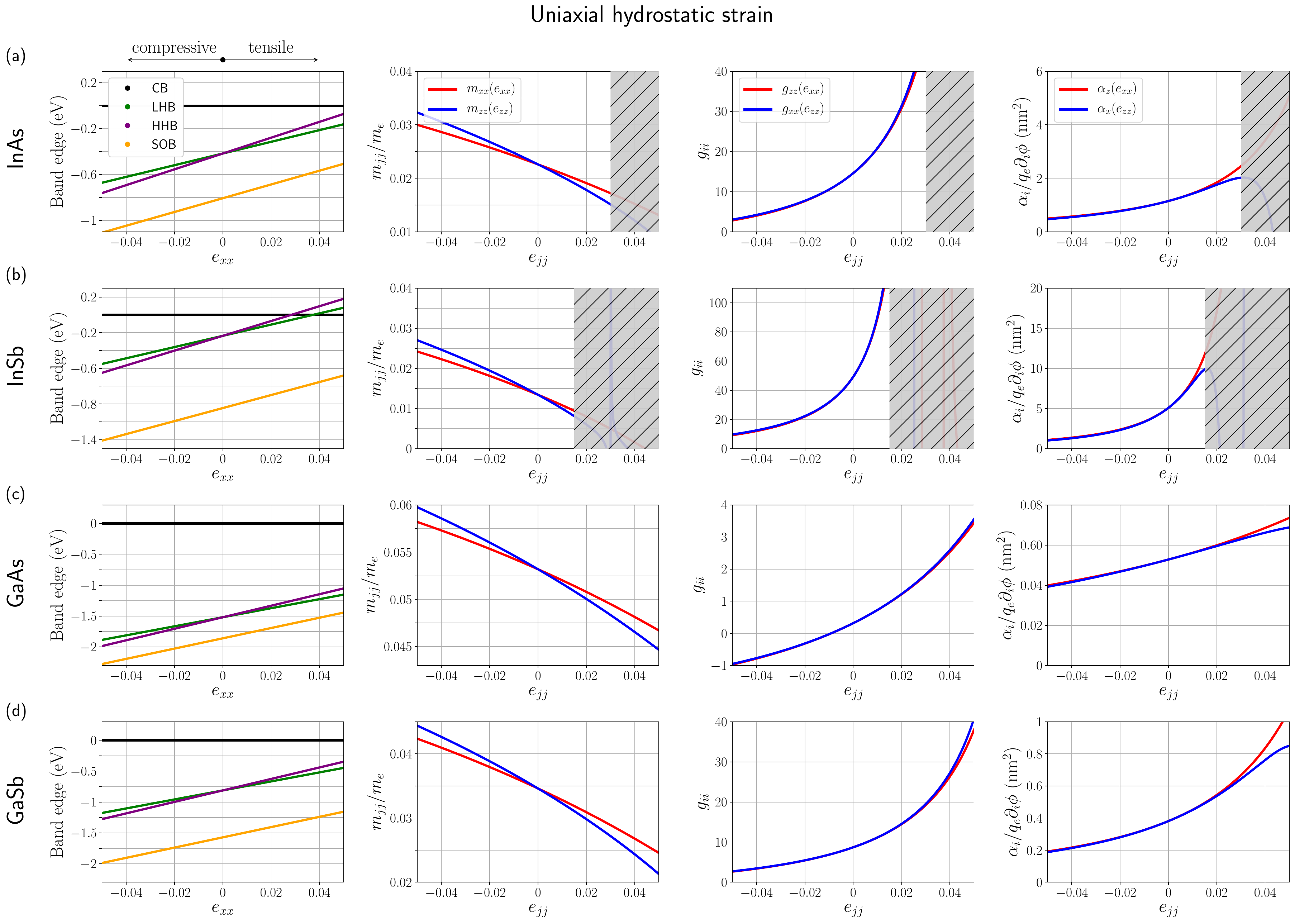}
    \caption{ \textbf{Uniaxial hydrostatic strain effective parameters.} Effective parameters obtained within the CB approximation under a uniaxial hydrostatic deformation. The panels display the band edge (first column), effective mass (second), $g$-factor (third), and SO coupling (fourth) for (a, first row) InAs, (b) InSb, (c) GaAs, and (d) GaSb. Gray regions indicate the strain values for which the CB approximation breaks down and the results are therefore unreliable. Different line colors denote different directions of applied deformation (see legends on top). A deformation along the $y$-direction is equivalent to one along $x$ due to the zincblende lattice symmetry. The parameters used for these simulations can be found in Appendix~\ref{AP:parameters}.}
    \label{FigSM1}
\end{figure}

\begin{figure}
    \centering
    \includegraphics[width=1\columnwidth]{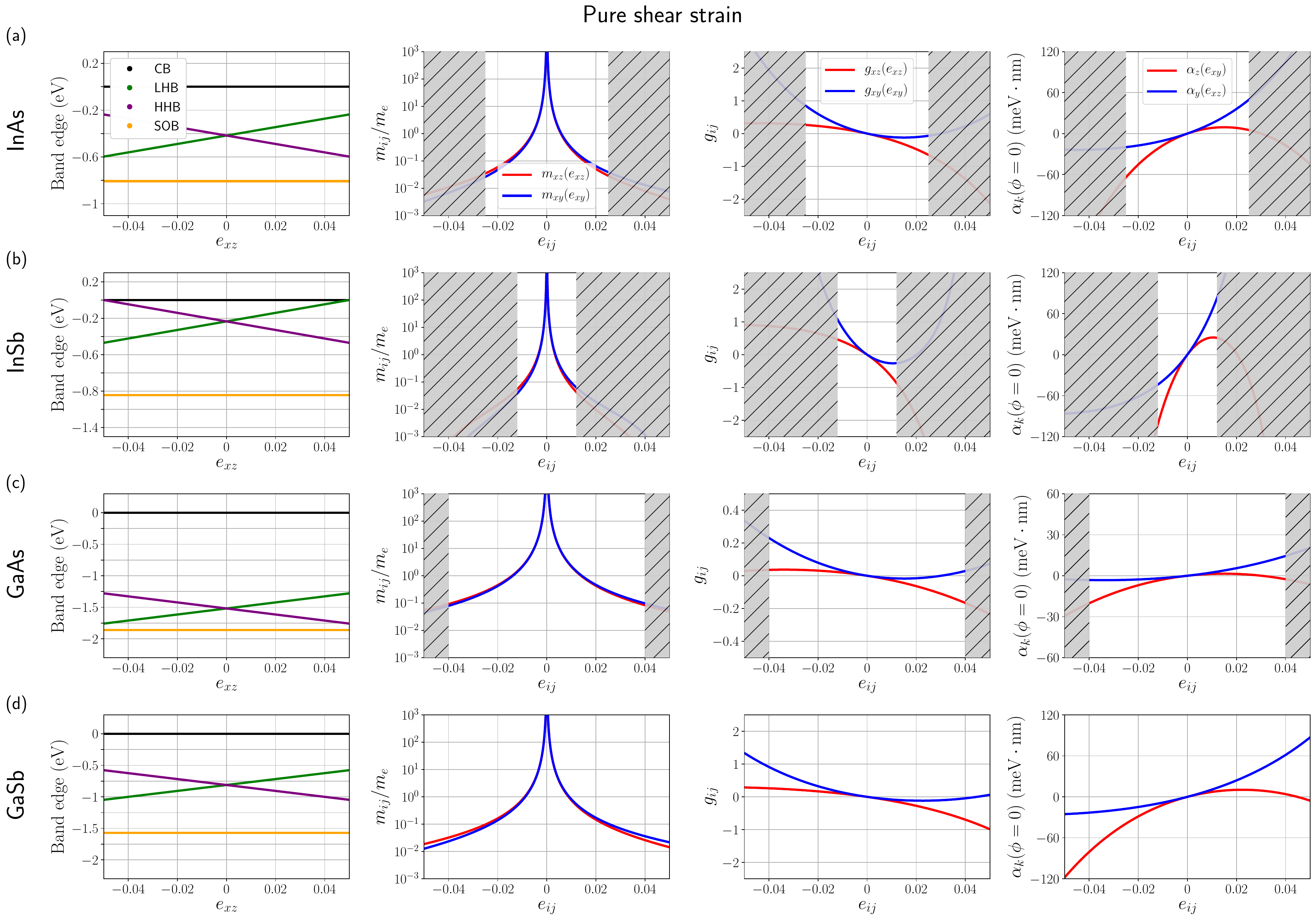}
    \caption{ \textbf{Pure shear strain effective parameters.} Same as in Fig.~\ref{FigSM1} but for a pure shear deformation.}
    \label{FigSM2}
\end{figure}

In Fig.~\ref{FigSM1} and Fig.~\ref{FigSM2}, we present the effective parameters obtained within the CB approximation under uniaxial hydrostatic deformation and pure shear deformation, respectively. Each row, (a)–(d), corresponds to a different material: (a) InAs, (b) InSb, (c) GaAs, and (d) GaSb. Each column shows a distinct parameter: the first column the band edges (CB, LHB, HHB, or SOB; see legend), the second the effective masses, the third the effective $g$-factor, and the fourth the SO coupling. Different line colors denote the direction along which the strain is applied (see legend). For example, in Fig.~\ref{FigSM1}, red lines correspond to a deformation along $x$, while blue lines correspond to a deformation along $z$. In this case, a deformation along $y$ has the same effect as along $x$, owing to the zincblende crystal symmetry. Gray regions indicate strain values for which the CB approximation breaks down, rendering the extracted effective parameters unreliable.

\section{Parameters for the simulations}
\label{AP:parameters}
In this Appendix, we list all the material parameters employed in our simulations, all of which are taken from Ref~\onlinecite{JAP:Vurgaftman01}. Table~\ref{tab:table_params} summarizes the parameters required for the 8B Kane model, which are the same as those used in the CB approximation. These parameters are used for Figs.~\ref{Fig1} to \ref{Fig3} for InAs, as well as in Figs.~\ref{FigSM1} and~\ref{FigSM2} for the full set of materials. For Fig.~\ref{Fig4}, in addition to the CB parameters for InAs, we also impose specific geometric constraints, which are summarized in Table~\ref{tab:table_params2}. These also include the dielectric permittivity of the different materials in the nanostructure and the charge accumulation that is typically present at the nanowire or 2DEG facets~\cite{Nano:Thelander10, AdvSci:Schuwalow21}.

\begin{table}
\centering
\caption{\label{tab:table_params}%
Material parameters employed in the 8B k$\cdot$p model at $T=0$~K, extracted from Ref.~\onlinecite{JAP:Vurgaftman01}. The same set of parameters is used consistently in all figures, both for the full 8B calculations and for the effective CB Hamiltonian. We fix the Fermi level at the CB edge.} 
\setlength\tabcolsep{0pt}
\begin{tabular*}{0.6\textwidth}{@{\extracolsep{\fill} } lcccc }
\hline\hline
\textrm{}&
\textrm{InAs}&
\textrm{InSb}&
\textrm{GaAs}&
\textrm{GaSb} \;\;\; \\
\hline\hline
$\Delta_{\rm g}$ [eV] & 0.417 & 0.235 & 1.519 & 0.812 \;\;\;\\
$\Delta_{\rm soff}$ [eV] & 0.390 & 0.810 & 0.341 & 0.760 \;\;\; \\
$P$ [eV$\cdot$nm] &  0.905 & 0.942 & 1.047 & 1.014 \;\;\;\\
$\gamma_{1}$ & 20 & 34.8 & 6.98 & 13.4 \;\;\; \\
$\gamma_{2}$  & 8.5 & 15.5 & 2.06 & 4.7 \;\;\; \\
$\gamma_{3}$  & 9.2 & 16.5 & 2.93 & 6.0 \;\;\; \\
$a$ [eV] & 6 & 7.3 & 8.33 & 8.3  \;\;\;\\ 
$b$ [eV] & 1.8 & 2 & 2 & 2 \;\;\; \\ 
$d_{\rm v}=d_{\rm cv}$ [eV]  & 3.6 & 4.7 & 4.8 & 4.7 \;\;\; \\  \hline\hline
\end{tabular*}
\end{table}

\begin{table}
\centering
\caption{\label{tab:table_params2}%
Parameters used in Fig.~\ref{Fig4}. The parameter $W_{\rm active}$ denotes the thickness of the active layer, taken as InAs in all cases, with dielectric permittivity $\epsilon_{\rm InAs} = 15.5$. The parameter $W_{\rm insulating}$ corresponds to the thickness of the insulating layer separating the active region from the gate, chosen as AlSb with dielectric permittivity $\epsilon_{\rm AlSb} = 10.9$. In Fig.~\ref{Fig4}(a), $R_{\rm core}$ specifies the radius of the inner insulating core. In all simulations, we include a surface charge density of $\rho_{\rm surf} = 0.05~\mathrm{e}/\mathrm{nm}^2$ at the nanowire facets (within the last $1$~nm near the interface) to account for the accumulated surface charge typically present in low-dimensional SMs~\cite{Nano:Thelander10, AdvSci:Schuwalow21}. The temperature is fixed to $T = 1.7$~K in all cases.}
\setlength\tabcolsep{0pt}
\begin{tabular*}{0.6\textwidth}{@{\extracolsep{\fill} } lcccc }
\hline\hline
\textrm{}&
\textrm{Fig.~\ref{Fig4}(a)}&
\textrm{Fig.~\ref{Fig4}(b)}&
\textrm{Fig.~\ref{Fig4}(c)}&
\textrm{Fig.~\ref{Fig4}(d)} \;\;\; \\
\hline\hline
$W_{\rm active}$ [nm] & 10 & 40 & 10 & 10 \;\;\;\\
$W_{\rm insulating}$ [nm] & 5 & 10 & 5 & 5 \;\;\; \\
$R_{\rm core}$ [nm] &  35 & - & - & - \;\;\;\\  \hline\hline
\end{tabular*}
\end{table}

\end{document}